\documentclass[amsmath,amssymb,noshowpacs,aip,pop,
preprint,
citeautoscript  
]{revtex4-1}
\usepackage[mathscr]{eucal}
\usepackage{amssymb}
\usepackage{amsmath}
\usepackage{dcolumn}
\usepackage{bm}
\usepackage{graphicx}
\usepackage{epstopdf}
\usepackage{subfigure}
\usepackage{color}
\usepackage{xcolor}
\usepackage{setspace}
\usepackage[percent]{overpic}
\usepackage{booktabs}
\usepackage{lineno}
\usepackage{amsthm}
\usepackage{siunitx}
\usepackage{geometry}
\usepackage{hyperref}
\newtheoremstyle{query}%
{}{}
{\color{red}}
{}
{\sffamily\bfseries}{:}{12pt}
{}
\theoremstyle{query}

\newcommand{\AzhixinA}[1]{\textcolor{black}{#1}}
\newcommand{\AlanaB}[1]{\textcolor{black}{#1}}

\newcommand{\lanaA}[1]{\textcolor{black}{#1}}
\newcommand{\zhixinA}[1]{\textcolor{black}{#1}}
\newcommand{\lanaB}[1]{\textcolor{black}{#1}}
\newcommand{\zhixinB}[1]{\textcolor{black}{#1}}
\newcommand{\lanaC}[1]{\textcolor{black}{#1}}
\newcommand{\zhixinC}[1]{\textcolor{black}{#1}}

\UseRawInputEncoding
\begin{document}

\title{Gyrokinetic simulations of neoclassical electron transport and bootstrap current generation in tokamak plasmas in the TRIMEG code}

\author{Lana Rekhviashvili} 
\affiliation{Max Planck Institut f\"ur Plasmaphysik, 85748, Garching, Germany}
\affiliation{Technische Universit\"at M\"unchen, 80333, Munich, Germany}

\author{Zhixin Lu}
\email{zhixin.lu@ipp.mpg.de}
\affiliation{Max Planck Institut f\"ur Plasmaphysik, 85748, Garching, Germany}

\author{Matthias Hoelzl}
\affiliation{Max Planck Institut f\"ur Plasmaphysik, 85748, Garching, Germany}

\author{Andreas Bergmann}
\affiliation{Max Planck Institut f\"ur Plasmaphysik, 85748, Garching, Germany}

\author{Philipp Lauber}
\affiliation{Max Planck Institut f\"ur Plasmaphysik, 85748, Garching, Germany}

\date{\today}

\begin{abstract}
    For magnetic confinement fusion in tokamak plasmas, some of the limitations to the particle and energy confinement times are caused by turbulence and collisions between particles in toroidal geometry, which determine the ``anomalous'' and the neoclassical transport, respectively. Neoclassical effects are also responsible for the intrinsically generated bootstrap current, and only the self-consistent modeling of neoclassical and turbulent processes can ultimately give accurate predictive results. In this work, we focus on the implementation of neoclassical physics in the gyrokinetic code TRIMEG, which is a TRIangular MEsh-based Gyrokinetic code that can handle both the closed and open field line geometries of a divertor tokamak. 
    We report on the implementation of a simplified Lorentz collision operator in TRIMEG. For comparison with neoclassical theory, the calculation of flux surface averages is necessary. Since the code uses an unstructured mesh, a procedure for calculating the flux surface averages of particle and energy fluxes and the bootstrap current is derived without relying on the poloidal coordinate, which is useful also for other simulations in unstructured meshes. With the newly implemented collision operator, we study electron transport and bootstrap current generation in a plasma with a  finite density gradient but uniform temperature for various simplified and realistic geometries. In the comparison to neoclassical theory, good agreement is obtained for the large aspect ratio case regarding the particle and energy fluxes as well as the bootstrap current. However, some discrepancies are observed at moderate aspect ratio and for a case with the realistic geometry of the ASDEX Upgrade tokamak. These deviations can be explained by different treatments and approximations in theory and simulation. In this paper, we demonstrate the capability to calculate the electron transport and bootstrap current generation in TRIMEG, which will allow for the self-consistent inclusion of neoclassical effects in gyrokinetic simulations in the future.
\end{abstract}

\maketitle

\section{Introduction}
\label{sec:introduction}
Neoclassical transport determines the minimum level of transport for plasma confinement in tokamaks or stellarators and is one of the key issues in the optimization of experimental devices \cite{beidler2021demonstration}.  The bifurcation of neoclassical transport for steep equilibrium profiles has also been proposed as a possible mechanism for the transition to the high confinement regime in tokamak plasmas \cite{helander1998bifurcated}. 
The neoclassical theory has been developed for the comparison and the prediction of the confinement properties in toroidally confined plasmas \cite{hinton1976theory,hirshman1981neoclassical}. Standard neoclassical theory, while illustrating the dominant physics mechanism, assumes a static magnetic equilibrium and a steady-state solution and solves the gyrokinetic/drift kinetic equation asymptotically using the poloidal Larmor radius as a small parameter. 
It is demonstrated from the gyrokinetic simulation that the large orbit effects can lead to significant discrepancy compared with traditional neoclassical theory \cite{lin1997large}.
It is also found that the nonlocal effect of neoclassical transport is important in matching experimental results \cite{wang2006nonlocal}.
Recently, neoclassical transport in the plasma edge with open field lines has attracted significant attention \cite{chang2004numerical}.
In edge transport studies, a fully nonlinear collision operator has been developed to handle the non-Maxwellian electron distribution function in the plasma edge \cite{zhao2019solps}.
In all these aspects, the traditional neoclassical theory is not applicable. Furthermore, the self-consistent interaction of neoclassical processes and turbulence is eventually needed to capture the full picture of dynamical processes in fusion devices and provide reliable predictions.

Particle simulations constitute a powerful tool for the studies of neoclassical physics \cite{lee1983gyrokinetic,lin1995gyrokinetic,wang2006nonlocal,bergmann2001guiding}. The neoclassical simulation can  also provide a more consistent initial condition for turbulence simulations with a neoclassical particle distribution \cite{wang2006gyro,vernay2010neoclassical}. In this work, we focus on the implementation of the particle scheme in an unstructured mesh, following our previous work of the TRIMEG (TRIangular MEsh based Gyrokinetic) code \cite{lu2019development}. \lanaB{ The intention of this paper is to demonstrate the capabilities of the TRIMEG code and benchmark the implementation of the collision operator.} The Monte-Carlo integration for the calculation of the particle/energy fluxes and bootstrap current is developed in unstructured meshes. In addition, the traditional local neoclassical theory is compared with the global particle simulations. While the eventual goal is the full treatment of multiple species with the neoclassical electric field, in this work, we focus on electron transport without coupling to the neoclassical electric field and ion dynamics. 

The remainder of this paper is organized as follows. 
In Sec.~\ref{sec:models}, the basic equations are presented. We start from general equations and their reduction to those that are implemented in the TRIMEG code. The gyrokinetic equations and the collision operator are listed. Specific issues such as the flux surface average and toroidal average in an unstructured mesh are discussed. 
In Sec.~\ref{sec:results}, three typical cases with increasing complexity are studied. The local theoretical results of particle/energy fluxes and bootstrap current are calculated and compared to those obtained from the gyrokinetic simulations. The agreement and the discrepancy between the local theory and the global particle simulations, as well as the reasons or the possible limitation of the local theory, are discussed. Conclusions as well as an outlook to future work are given in Sec.~\ref{sec:conclusions}.

\section{Models and Equations}
\label{sec:models}
\subsection{General equations in $\delta f$ simulations}
\label{sec:new-developments-trimeg}
Using the $\delta f$ scheme, the distribution function $f$ is
split into a fixed part $f_0$
and a perturbation $\delta\! f$,
namely, $f=f_0+\delta f$.
$f_0$ is a known distribution and $\delta f$ is obtained from 
\begin{align}
\label{eq:ddfdt}
    &\frac{\rm{d}}{{\rm d}t}\delta f(t)  = -  \frac{{\rm d}}{{\rm d}t} f_0 +C(f,f) \;\;, \\
\label{eq:ddt_operator}
    &\frac{{\rm d}}{{\rm d}t} = \frac{\partial}{\partial t} + {\bf\dot R}\cdot\nabla + \dot v_\|\frac{\partial}{\partial v_\|} \;\;,
\end{align}
where $C(f,f)$ is the collision operator, ${\bf R},v_\|,\mu$ are the guiding center coordinates, \lanaA{the magnetic moment devided by the particle mass is adopted as one of the phase space coordinates defined as} $\mu=v_\perp^2/(2B)$, $v_\|$ and $v_\perp$ are the velocities along and perpendicular to the magnetic field, respectively. 

In this work, the Maxwell distribution is chosen ($f_0=f_{\rm{M}}$), 
\begin{eqnarray}
 f_{\rm{M}}=\frac{n_0}{(2T/m)^3\pi^{3/2}} \exp\left\{-\frac{mv_\|^2}{2T}-\frac{m\mu B}{T}\right\} \;\;,
\end{eqnarray}
where $T$ and $n$ are functions of the coordinates $(R,Z)$ in the poloidal cross-section, and thus
\begin{eqnarray}\label{eq:dlnfMdt_all}
 \frac{{\rm d}}{{\rm d}t}\ln f_{\rm{M}} 
 &=& ({\bf v}_\|+{\bf\dot R}_d+\delta {\bf\dot R} ) \cdot 
 \left[
 {\vec\kappa}_n + \left(\frac{mv_\|^2}{2T}+\frac{m\mu B}{T}-\frac{3}{2}\right)\vec\kappa_T 
 -\frac{m\mu B}{T}\vec\kappa_B
 \right] \nonumber\\
 &-& (\dot v_{\parallel,0} + \delta \dot v_{\|}  ) \frac{mv_\|}{T} \;\;, 
\end{eqnarray}
where $\vec\kappa_{n,T,B}\equiv \nabla\ln \{n,T,B\}$, ${\bf\dot R}_d$  is the magnetic drift velocity, $\delta {\bf\dot R}$ is the perturbed velocity due to the wave field and the neoclassical electric field, $\dot v_{\parallel,0}$ and $\delta \dot v_{\|}$ are due to the mirror force and the wave/neoclassical fields, respectively. 
Furthermore, the perturbed parts $\delta \Dot{\mathbf{R}}$ and $\delta \Dot{v_\parallel}$ contain contributions from the turbulence as well as the neoclassical electric field
and can be represented as \lanaA{ $\delta \Dot{\mathbf{R}}=\delta \Dot{\mathbf{R}}_{turb} + \delta \Dot{\mathbf{R}}_{E_{ZF}} + \delta \Dot{\mathbf{R}}_{E_{nc}}$, where $\delta \Dot{\mathbf{R}}_{turb} $ is the non axi-symmetric turbulence component while $\delta \Dot{\mathbf{R}}_{E_{ZF}}$ and  $\delta \Dot{\mathbf{R}}_{E_{nc}}$  are related the $n=0$ axi-symmetric field due to turbulence and collisions between particles, respectively. }
For models with \lanaA{$\Phi\AzhixinA{_{nc}}=\Phi\AzhixinA{_{nc}}(r)$, where $\Phi$ is the scalar potential and}  the radial-like coordinate $r$ is a function of the magnetic flux, we readily have $\delta \dot v_\parallel=0$, which is the assumption used in some neoclassical studies \cite{hinton1976theory}. More generally, the neoclassical \lanaA{scalar potential}  can vary in the poloidal direction and it is denoted as \lanaA{$\Phi_{nc}=\Phi_{nc}(R,Z)$}.
In the previous TRIMEG work \cite{lu2019development,lu2021development,Lu_2023}, neoclassical effects were ignored.
For the neoclassical studies in this work, we ignore the $n\ne0$ modes and we have
\begin{eqnarray}\label{eq:dlnfMdt_nc}
 \frac{{\rm d}}{{\rm d}t}\ln f_{\rm{M}} 
 &=& ({\bf v}_\|+{\bf\dot R}_d+\delta {\bf\dot R}_{E_{nc}} ) \cdot\left[
 {\vec\kappa}_n + \left(\frac{mv_\|^2}{2T}+\frac{m\mu B}{T}-\frac{3}{2}\right)\vec\kappa_T
 -\frac{m\mu B}{T}\vec\kappa_B
 \right] \nonumber\\
 &-& (\dot v_{\parallel,0} + \delta \dot v_{\| E_{nc}}  ) \frac{mv_\|}{T} \;\;,
\end{eqnarray}
where $\delta\dot{\mathrm{R}}_{E_{nc}}$ and $\delta \dot v_{\| E_{nc}}$ are the perturbed velocity and  acceleration due to the neoclassical electric field, respectively. In this work, since we focus on electron transport, $E_{nc}$ is not taken into account, as adopted by previous work \cite{lin1996gyrokinetic}. 
\lanaA{In addition, since the electron Larmor radius is negligible compared to the scale length of the equilibrium gradient, the drift kinetic equations are solved. }
Note that Eqs. \ref{eq:dlnfMdt_all}--\ref{eq:dlnfMdt_nc} are written in $({\bf R},v_\|,\mu)$ coordinates, for demonstrating the general form of the models without and with neoclassical physics. The right-hand side can be also written in  $({\bf R},E,\mu)$ coordinates, in order to use the constants of motion $(E,\mu)$, where $E$ is the energy. Further simplifications, e.g. omission of the electric field, is adopted to get the equation implemented in TRIMEG in this work, as shown in Eq. \ref{eq:dwdt_vdkappa} in the next section. 

\subsection{Discretization of the distribution function}
When representing the distribution function in simulations, we need to discretize it.
In the $\delta f$ scheme, $f_0$ and $\delta f$ can be represented as follows with respective weight fields $P(z,t)$ and $W(z,t)$ \cite{lanti2019global}
:
\begin{eqnarray}
\label{eq:f0from_marker}
    f_0(z,t) = \frac{N_{ph}}{N}P(z,t)g(z,t) \approx \frac{N_{ph}}{N}\sum_{i=1}^{N}p_i\frac{\delta(z-z_i(t))}{J(z)}\;\;,
    \\
\label{eq:f1from_marker}
    \delta f(z,t) = \frac{N_{ph}}{N}W(z,t)g(z,t) \approx \frac{N_{ph}}{N}\sum_{i=1}^{N}w_i\frac{\delta(z-z_i(t))}{J(z)}\;\;,
\end{eqnarray}
where $N_{ph}$ and $N$ are the physics particle number and marker number, respectively, $z$ denotes the phase space coordinate $({\bf R},v_\|,\mu)$, whose Jacobian is $J(z)$.
Furthermore, the time evolution for the weights in the collisionless case can be written as
\begin{eqnarray}
    &&\frac{d}{dt} w_i(t)=-p_i\frac{d}{dt}\ln f_0(z_i(t))\;\;, \\
    &&\frac{d}{dt} p_i(t)=p_i\frac{d}{dt}\ln f_0(z_i(t))\;\;, \\
    &&\frac{d}{dt}(w_i(t)+p_i(t))=0\;\;.
\end{eqnarray}
Assuming that the background distribution is of zeroth order in $\rho_p / R_0$, where $\rho_p$ is the poloidal gyro-radius, the time evolution for the weight equation can be written as \cite{lin1996gyrokinetic}
\begin{equation}
\label{eq:weight-eq}
    \frac{dw}{dt}=-(1-w)\Vec{v}_d\cdot\nabla \ln{f_0}.
\end{equation}
To take into account the collision effects, at each time step, the collision operators can be applied after the collisionless dynamics \cite{lanti2019global}. Theoretically, the collision operator is taken into account in particle simulations by solving the Langevin equation, as we will discuss in the following section. 

\subsection{Collision operator}
\lanaA{Realistic non-liniear collision operators have been implemented in the XGC code recently \cite{hager2016gyrokinetic}. However, in our studies for simplicity we use the linearized collision operator.}
By treating the heavy species as infinitely massive with negligible thermal velocity, the collision operator becomes much simpler and the Lorentz collision operator is readily obtained \cite{hazeltine2019framework}
\begin{eqnarray}
\label{eq:lorentz-col-op}
    C_{ei}(\delta f_e) = \nu_0 \frac{1}{2} \frac{\partial}{\partial \xi}(1-\xi^2) \frac{\partial}{\partial \xi}\;\;.
\end{eqnarray}
In this work, only the electron-ion Lorentz collision operator is used, given by Eq. \ref{eq:lorentz-col-op}, which is written using the Monte-Carlo method as \cite{lin1996gyrokinetic}
\begin{equation}
\label{eq:descretized-col-op}
    \xi = \xi_0 (1-\nu \Delta t) + (\lanaA{\mathcal{R}}-0.5)[12(1-\xi_0^2)\nu \Delta t]^{1/2},
\end{equation}
where $\xi=v_\parallel/v$, $\nu$ is the collision frequency, and \lanaA{$\mathcal{R}$} is a uniform random number between 0 and 1 \cite{lin1996gyrokinetic}. \zhixinB{Since the Lorentz model is valid in the limit of large ion-to-electron mass ratio, the focus of this work is the benchmark of the code with the theoretical results in the simplified case  while more realistic collision operator needs to be implemented in future in order to compare with experiments.}
\subsection{Implemented Equations and normalizations}
For our neoclassical studies, the TRIMEG code was modified as described in this section. In addition to the discussions in \ref{sec:new-developments-trimeg}, the $\delta f$ model was adopted and a weight equation was implemented as given by Eq. \ref{eq:weight-eq} in $(R,Z,\phi)$ coordinates
\begin{equation}\label{eq:dwdt_vdkappa}
    \frac{dw}{dt}=(1-w)(\Vec{v}_d\cdot\Vec{\kappa})=(1-w)(v_{d,R}\kappa_R+v_{d,Z}\kappa_Z),
\end{equation}
where $\kappa$ is the negative gradient of the Logarithm of  the equilibrium distribution function given by
\begin{equation}
    \begin{split}
    \Vec{\kappa}&=-\frac{\partial  \ln f_0}{\partial \Vec{x}}=-\left(\frac{\partial  \ln f_0}{\partial n}\frac{\partial n}{\partial R}+\frac{\partial  \ln f_0}{\partial T}\frac{\partial T}{\partial R}\right)\hat{R}-\left(\frac{\partial  \ln f_0}{\partial n}\frac{\partial n}{\partial Z}+\frac{\partial  \ln f_0}{\partial T}\frac{\partial T}{\partial Z}\right)\hat{Z}\\
    &=-\left[\frac{\partial  \ln n}{\partial R}+\left(\frac{mv^2}{2T}-\frac{3}{2}\right)\frac{\partial  \ln T}{\partial R}\right]\hat{R}-\left[\frac{\partial  \ln n}{\partial Z}+\left(\frac{mv^2}{2T}-\frac{3}{2}\right)\frac{\partial  \ln T}{\partial Z}\right]\hat{Z}.
    \end{split}
\end{equation}
Taking into account the normalization given in table \ref{tab:trimeg-normalization}, the normalized equation implemented in the code is
\begin{equation}
    \begin{split}
        \frac{dw}{dt}=(1-w)\left\{-\Bar{v}_{d,R}\left[\frac{\partial \ln \lanaB{\bar{n}}}{\partial \lanaB{\bar{R}}}+\left(\frac{\Bar{m}\Bar{v}_\parallel^2}{\Bar{T}}+\frac{2\Bar{\mu}\Bar{m}\Bar{B}}{\Bar{T}}-\frac{3}{2}\right)\frac{\partial  \ln \lanaB{\bar{T}}}{\partial \lanaB{\bar{R}}}\right] \right. \\ 
        \left.-\Bar{v}_{d,Z}\left[\frac{\partial \ln \lanaB{\bar{n}}}{\partial \lanaB{\bar{Z}}}+\left(\frac{\Bar{m}\Bar{v}_\parallel^2}{\Bar{T}}+\frac{2\Bar{\mu}\Bar{m}\Bar{B}}{\Bar{T}}-\frac{3}{2}\right)\frac{\partial  \ln \lanaB{\bar{T}}}{\partial \lanaB{\bar{Z}}}\right] \right\}.
    \end{split}
\end{equation}

\begin{table}[h]
    \centering
    \begin{tabular}{|r||l|}
         \hline
        \multicolumn{2}{|c|}{TRIMEG Normalization units} \\
        \hline
        $m_N$& $m_i$  \\[0.1cm]
        $R_N$& $1$ m  \\[0.1cm]
        $T_N$& $T_{i,ref}$  \\
        $v_N$& $v_{th,i}=\sqrt{\frac{2T_N}{m_N}}$  \\[0.1cm]
        $t_N$& $\frac{R_N}{v_N}$  \\[0.1cm]
        $B_N$& $1$ Tesla  \\[0.1cm]
        \hline
    \end{tabular}
    \caption{Normalization used in TRIMEG.}
    \label{tab:trimeg-normalization}
\end{table}

\begin{figure}[h]
    \centering
    \includegraphics[width=0.48\textwidth]{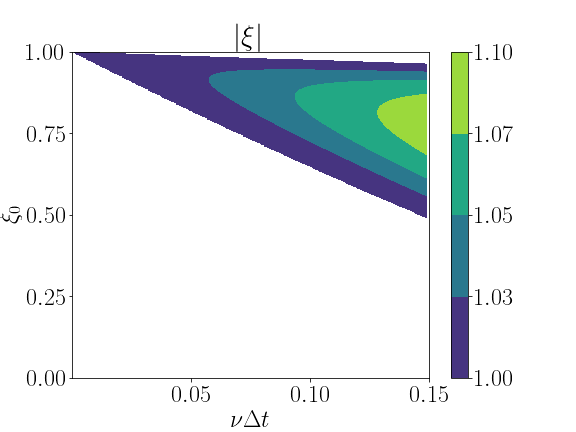}
    \caption{Values of $\xi_0$ and $\nu\,\Delta t$ when $\xi$ becomes larger than 1, during evolution using the Lorentz collision operator given in Eq. \ref{eq:descretized-col-op}. Here the random number \lanaA{$\mathcal{R}$} is taken to be $0.9$, and $\nu$ represents the collision frequency.}
    \label{fig:collision-operator}
\end{figure}

To take into account the collisional effects, the discretized Lorentz collision operator given by Eq. \ref{eq:descretized-col-op} is implemented in the code,
\begin{equation}
    \xi = \xi_0 (1-\Bar{\nu} \Delta \lanaB{\bar{t}}) + (\lanaA{\mathcal{R}}-0.5)[12(1-\xi_0^2)\Bar{\nu} \Delta \lanaB{\bar{t}}]^{1/2},
\end{equation}
where $\xi$ is the particle pitch as defined earlier, $\Bar{\nu}=\nu\,t_N$ is the normalized collision frequency \lanaB{and $\bar{t}$ is the normalized time}. Furthermore, regardless of the choice of $\lanaB{\nu}\,\Delta t$, due to the choice of \lanaA{$\mathcal{R}$} being random, $|\xi|$ can become greater than 1, as shown in Fig. \ref{fig:collision-operator}. This would cause a nonphysical solution with $|v_\|/v|>1$, which was fixed by re-setting $|\xi|$ equal to exactly 1 in those cases. \lanaB{Frequent re-setting of the pitch would break the uniformity of the random distribution due to over-sampling of $1,-1$. Hence, we want to avoid it by limiting the number of times the pitch becomes greater than 1.} For large values of $\lanaB{\nu}\,\Delta t (>0.1{\zhixinB 5})$, more markers end up with $|\xi|>1$ after the collision, and the re-setting operation is needed more frequently. For small values of $\lanaB{\nu}\,\Delta t (<0.1\lanaB{5})$, only a small portion of markers enter the $|\xi|>1$ zone, and the re-setting operation is needed less frequently. In the simulations, the value for $\lanaB{\nu}\,\Delta t$ is chosen to be be small enough ($\lanaB{\nu}\,\Delta t\lanaB{<0.15}$) to avoid frequent use of re-setting.
\zhixinB{While in Fig. \ref{fig:collision-operator}, the random number is close to $1$ ($\mathcal{R}=0.9$), the total portion of the $|\xi|>1$ particles is much smaller than $50\%$ for $\nu\Delta t=0.15$ since $\mathcal{R}\in[0,1)$. }

\subsection{Diagnosis for axisymmetric components in TRIMEG}
\label{sec:flux-surface-average}
In the studies of neoclassical transport, the particle and energy fluxes and the bootstrap current are all axisymmetric variables and need to be calculated numerically.
An important routine implemented in the code during our studies is the flux surface average calculation. For any function $F$, the flux surface average is \cite{hinton1976theory}
\begin{equation}
\label{eq:flux-surf-average}
    \langle F \rangle = \frac{\int_{\Delta V} d^3\Vec{R} F}{\int_{\Delta V} d^3\Vec{R}} = \frac{d\psi}{dV}\int \frac{dS}{|\nabla\psi|}F,
\end{equation}
where $\Delta V$ is the small volume between two adjacent flux surfaces, and $dS$ is the area element on the flux surface. 
For general geometry, we can substitute $\delta f$ with Eq. \ref{eq:f1from_marker} and re-write the flux surface average given by Eq. \ref{eq:flux-surf-average} as
\begin{equation}
    \langle F \rangle=\frac{1}{\Delta V} \int_{\Delta {V}} d^3 \Vec{R}' \frac{1}{\Delta \hat{V}}\int_{\Delta\hat V} d^3\Vec{R} \frac{1}{J_{\Vec{R}}}\frac{N_{ph}}{N_m} \sum^{N_m}_{p=1}w_p\delta(\Vec{R}-\Vec{R}_p)\delta(\Vec{v}-\Vec{v}_p)X\;\;.
\end{equation}
After taking the integrals this is simplified to
\begin{equation}
\label{eq:Fsurfavg}
    \langle F \rangle=\frac{1}{\Delta V}\frac{N_{ph}}{N_m} \sum_{p\in\Delta V}w_p X_p.
\end{equation}
\lanaA{We can also readily calculate the flux surface average for 3D variables, namely,}
\begin{equation}
\label{eq:F3d}
    F(\psi,\theta,\phi)=\frac{\int_{\Delta\Tilde{ V}}d^3\Vec{R} \delta f X}{\int_{\Delta\Tilde{ V}}d^3\Vec{R}},
\end{equation}
\lanaA{where $\psi$ is the poloidal magnetic flux, $\theta$ is a poloidal-like coordinate, and} $\Delta\Tilde{V}$ is an infinitesimal volume. The flux surface average of $F$ yields the same result in Eq. \ref{eq:Fsurfavg}.

In some cases, the toroidal average is used for the analysis instead of the flux surface average (which is relevant for future applications with asymmetric poloidal structures): 
\begin{equation}
    \langle F \rangle_{\phi}= \frac{\int_0^{2 \pi} d\phi F}{\int_0^{2 \pi} d\phi}.
\end{equation}
When plugging in the definitions for $F$, the average can be written as
\begin{equation}
    \langle F \rangle_{\phi}= \frac{1}{\int_0^{2 \pi} d\phi}\frac{1}{\Delta \Tilde{V}}\frac{N_{ph}}{N_m}\int_{\Delta \Tilde{V}}d^3\Vec{R} \int d\phi\sum^{N_m}_{p=1}w_p\delta(\Vec{R}-\Vec{R}_p)\delta(\Vec{v}-\Vec{v}_p)X,
\end{equation}
where if integrals over the toroidal direction and over an arbitrarily small volume are taken, this equation can be rewritten as 
\begin{equation}
    \langle F \rangle_{\phi}= \frac{1}{2\pi \Delta \Tilde{S} }\Delta \phi \frac{N_{ph}}{N_m}\sum_{p\in \Delta \Tilde{S}}\frac{1}{R_p}w_p X_p,
\end{equation}
where $\Delta\tilde{V}=\tilde R\Delta\tilde{S}\Delta\phi$ has been adopted.

In our analysis, we deal with the toroidal average or the flux surface average normalized by average density. Hence, if we take into account that $N_{ph}=V_{tot} \langle n \rangle_{V}$, where $V_{tot}$ is the total volume and $\langle n \rangle_V$ is the average density in this volume, the toroidal average equation can be rewritten as 
\begin{equation}
    \frac{\langle F \rangle_{\phi}}{\langle n \rangle_V}=\frac{1}{2\pi \Delta \Tilde{S}}\frac{V_{tot}}{N_{m}}\sum_{p\in\Delta \Tilde{S}}\frac{1}{{R}_p}w_p X_p,
\end{equation}
where $\Delta \Tilde{V}=\Delta \Tilde{S} \Tilde{R} \Delta \phi$ was used.
The flux surface average Eq. \ref{eq:Fsurfavg} can be also expressed in the same way. For the calculation of the annulus area in shaped tokamak geometry, the Monte-Carlo integration method is used and the annulus area is calculated at the beginning of the simulation when the markers are distributed uniformly.

\subsection{Diagnosis and benchmark using the local neoclassical transport theory}
\label{sec:equations-for-theory-in-code}
The benchmark is done by comparing the simulation results to the theoretical local electron transport model \cite{hinton1976theory,lin1996gyrokinetic}.
We can also write simplified equations for electrons in a circular geometry, where ion charge is equal to one, and assuming constant pressure and temperature for the ion species. From the discussions in previous work \cite{hinton1976theory}, we can summarize these equations as follows
\begin{eqnarray}
    A'_{1e}&=&\frac{\partial \ln{n_e}}{\partial r} - \frac{3}{2}\frac{\partial \ln{T_e}}{\partial r}, \\
    \Gamma_e &=& - n_e \frac{4}{3\sqrt{\pi}} \nu_{ei} \epsilon^{-3/2}\rho^2 q^2 \left( K_{11}A'_{1e}+K_{12}\frac{\partial \ln{T_e}}{\partial r} \right), \\
    q_e&=& - n_e T_e \frac{4}{3\sqrt{\pi}}\nu_{ei} \epsilon^{-3/2}\rho^2 q^2 
    \left(K_{12}A'_{1e}+K_{22}\frac{\partial \ln{T_e}}{\partial r}\right) \nonumber\\
    &-&\frac{5}{2}T_e \Gamma_e, \\
    \left< (J_\parallel-J_{\parallel s})/h \right> &=& - n_e \frac{T_e}{m_e v_{th,e}} e \epsilon^{-1/2} \rho q \left( K_{13} A'_{1e} + K_{23} \frac{\partial \ln{T_e}}{\partial r} \right),
\end{eqnarray}
where the dimensionless $K$ coefficients are fitted analytically using values for the numerical coefficients\cite{hinton1976theory} for charge number equal to 1 with $\nu_\ast=\sqrt{2}\nu q R/(\epsilon^{3/2}v_{\rm th})=3\sqrt{\pi}\nu_{*e}/4$ :
\begin{eqnarray}
\label{coeff-Hinton1}
    K_{11}=1.04\left( \frac{1}{1+2.01\nu_{*e}^{1/2}+1.53\nu_{*e}} + \frac{\epsilon^{3/2}(0.89^2/1.53)\nu_{*e}\epsilon^{3/2}}{1+0.89\nu_{*e}\epsilon^{3/2}} \right),\\
    K_{12}=1.20\left( \frac{1}{1+0.76\nu_{*e}^{1/2}+0.67\nu_{*e}} + \frac{\epsilon^{3/2}(0.56^2/0.67)\nu_{*e}\epsilon^{3/2}}{1+0.56\nu_{*e}\epsilon^{3/2}} \right),\\
    K_{22}=2.55\left( \frac{1}{1+0.45\nu_{*e}^{1/2}+0.43\nu_{*e}} + \frac{\epsilon^{3/2}(0.43^2/0.43)\nu_{*e}\epsilon^{3/2}}{1+0.43\nu_{*e}\epsilon^{3/2}} \right), \\
    K_{13}=2.3[1+1.02\nu^{1/2}_{*e}+1.07\nu_{*e}]^{-1}[1+1.07\nu_{*e}\epsilon^{3/2}]^{-1},\\
    K_{23}=4.19[1+0.57\nu^{1/2}_{*e}+0.61\nu_{*e}]^{-1}[1+0.61\nu_{*e}\epsilon^{3/2}]^{-1},\\
    K_{33}=1.83[1+0.68\nu^{1/2}_{*e}+0.32\nu_{*e}]^{-1}[1+0.66\nu_{*e}\epsilon^{3/2}]^{-1}.
\label{coeff-Hinton2}
\end{eqnarray}
The normalized equations used in diagnostics are
\begin{equation}
    \Bar{A}'_{1e}=R_N A'_{1e}=\frac{\partial A_{1e}}{\partial \Bar{r}},
\end{equation}
\begin{equation}
    \Bar{\Gamma}_e=\frac{\Gamma_e}{n v_N} = - \frac{4}{3\sqrt{\pi}} \Bar{\nu}_{ei} \epsilon^{-3/2}\Bar{\rho}^2 q^2 \left( K_{11}\Bar{A}'_{1e}+K_{12}\frac{\partial \ln{\lanaB{\bar{T}_e}}}{\partial \Bar{r}} \right),
\end{equation}
\begin{equation}
    \Bar{q}_e=\frac{q_e}{n v_N m_N v^2_N}= - \frac{1}{2} \frac{4}{3\sqrt{\pi}}\Bar{\nu}_{ei} \epsilon^{-3/2}\Bar{\rho}^2 q^2\left(K_{12}\Bar{A}'_{1e}+K_{22}\frac{\partial \ln{\lanaB{\bar{T}_e}}}{\partial \Bar{r}}\right) 
    -\frac{5}{4} \Bar{\Gamma}_e,
\end{equation}
\begin{equation}
    \label{eq:finite-normalized-jb}
    \Bar{j}_b=\frac{j_b}{e n v_N} = \frac{1}{2}\sqrt{\frac{m_i}{m_e}} \epsilon^{-1/2} \Bar{\rho} q \left( K_{13}\Bar{A}'_{1e} + K_{23}\frac{\partial \ln{\lanaB{\bar{T}_e}}}{\partial \bar{r}} \right),
\end{equation}
where the normalized quantities are denoted by a bar over the variables, and $\left< (J_\parallel-J_{\parallel s})/h \right> \equiv j_b$.
In the asymptotic limits of collisionality, these formulae can also be rewritten as described in the paper by Lin \cite{lin1996gyrokinetic}.
For the banana regime, in the limit of $\nu \rightarrow 0$, the following analytical equations for particle flux $\Gamma$, energy flux $Q$, and the bootstrap current $j_b$ can be obtained \cite[]{lin1996gyrokinetic}:
\begin{equation}
    \begin{split}
        &\Gamma= \left\langle\int d^3v v_{dr} f_1\right\rangle = \frac{3}{8} I_1 \nu \rho^2 \frac{q^2}{\epsilon^2}n(\kappa_n+\kappa_t), \\
        &Q= \left\langle\int d^3v \frac{1}{2}m v^2 v_{dr} f_1\right\rangle = \frac{5T}{2} \left(\Gamma+\frac{3}{8}I_1\nu\rho^2\frac{q^2}{\epsilon^2}n\kappa_T\right), \\
        &j_b= \left\langle \int d^3v \frac{v_{\parallel}}{h} f_1\right\rangle=\frac{3}{4}I_3\frac{c}{B_{p0}}\frac{dp}{dr},\\
    \end{split}
\end{equation}
where  $\langle...\rangle\equiv\int_0^{2\pi} h {d \theta}/{(2\pi)}$ represents the flux surface average. Additionally, $\epsilon=r/R_0$, $h\equiv1+\epsilon \cos \theta$, $\rho=mv_{th,e}c / (eB_0)$, $q=rB_0/(R_0B_{p0})$, $p=(3/2)nT$, and to the lowest order in $\epsilon$,
\begin{equation}
    I_1=I_3=1.38 \sqrt{2\epsilon}.
\end{equation}
For the collisional limit $\nu q R/v_{\rm th} \gg 1$, the following analytical solutions are obtained \cite{lin1996gyrokinetic}
\begin{equation}
    \begin{split}
        & \Gamma=\nu q^2 \rho^2 n (\kappa_n+\kappa_T),\\
        &Q=\frac{5T}{2}(\Gamma+\nu q^2 \rho^2 n \kappa_T). \\
    \end{split}
\end{equation}
The normalized form of the fluxes and bootstrap current are as follows, for the banana regime,
\begin{equation}
\label{eq:banana-normalized-Gamma}
    \Bar{\Gamma} = \frac{\Gamma}{n v_N}=\frac{3}{8}I_1\Bar{\nu}\Bar{\rho}^2\frac{q^2}{\epsilon^2}(\Bar{\kappa}_n+\bar{\kappa}_T),
\end{equation}
\begin{equation}
\label{eq:banana-normalized-Q}
    \Bar{Q} = \frac{Q}{n v_N m_N v_N^2}=\frac{5}{2}\lanaB{2\,\bar{T}_e} \left(\Bar{\Gamma} +\frac{3}{8}I_1\Bar{\nu}\Bar{\rho}^2\frac{q^2}{\epsilon^2}\Bar{\kappa}_T  \right),
\end{equation}
\begin{equation}
\label{eq:banana-normalized-jb}
    \Bar{j_b} = \frac{j_b}{en v_N}=\frac{9}{16}\sqrt{\lanaB{\frac{1}{\bar{m}_e}}}I_3\bar{\rho}\frac{q}{\epsilon}(\Bar{\kappa}_T+\Bar{\kappa}_n),
\end{equation}
where $\Bar{\kappa}_{(n,T)} = R_N\kappa_{(n,T)}$. For the collisional regime,
\begin{equation}
\label{eq:col-normalized-Gamma}
\Bar{\Gamma}=\Bar{\nu}q^2\Bar{\rho}^2(\Bar{\kappa}_n+\Bar{\kappa}_T),
\end{equation}
\begin{equation}
\label{eq:col-normalized-Q}
    \Bar{Q}=\frac{5}{2}\lanaB{2\,\bar{T}_e}\left( \Bar{\Gamma}+\Bar{\nu}q^2\Bar{\rho}^2\Bar{\kappa}_T\right).
\end{equation}
Comparisons to the simulation results are done by calculating values using the flux surface average described in section \ref{sec:flux-surface-average} during the simulation, where the values for $X$ are chosen as $X=v_r mv^2/2$ for the energy flux, and $X=v_r$ for the particle flux, where $v_r=\vec{v}\cdot\nabla\psi/|\nabla\psi|$ is the radial drift velocity. 
The analytical formulas described in this chapter are implemented 
in the Fortran code, where the values for $\kappa_n$ and $q$ are interpolated from the density profile and the magnetic equilibrium. These results are described in sections \ref{sec:results}.

\section{Simulation setup and results}
\label{sec:results}

\subsection{Electron transport results for the larger aspect ratio case}
\label{subsec:itpa_results}
In this section, we will consider the International Tokamak Physics Activity (ITPA) case, which has been defined in the benchmark of the Toroidal Alfv\'en Eigenmode (TAE) driven by energetic particles (EPs) \cite{konies2018benchmark}.
This is a Tokamak plasma with large aspect ratio ($a/R_0=0.1)$ and  concentric circular magnetic surfaces. The on-axis magnetic field is $3$ Tesla. The major radius $R_0=10$ m. The nominal safety factor is $q=1.71+0.16(r/a)^2$ with low magnetic shear. In generating the EQDSK file for the ad-hoc equilibrium, the analytical form 
\begin{eqnarray}
\label{eq:qbar1d}
    \Bar{q}=\Bar{q}_0+\Bar{q}_2 r^2\;\;,
\end{eqnarray}
is adopted, where $\Bar{q}=q\sqrt{1-r^2/R_0^2}$. A simplified match to the $q$ profile is used by letting 
\begin{eqnarray}
    \Bar{q}_0=q_0=1.71\;\;, \;\;
    \Bar{q}_2=q_2=0.16\;\;,
\end{eqnarray} which is a good approximation for moderate or large aspect ratios. 
In our study of neoclassical transport and bootstrap current generation, we only keep the electrons as the kinetic species and assume a uniform electron temperature. The electron density gradient is nonuniform and the following profile is used for solving the weight equation \cite{lu2021development}
\begin{eqnarray}
\label{eq:itpa-density-profiles}
	n_{\rm{e}}(r)=n_{\rm{e},0}c_3 \exp\left( -\frac{c_2}{c_1} \tanh\frac{r-c_0}{c_2}\right)\;\;, \\
\frac{d\ln n_e}{dr} = -\frac{1}{c_1} \left[ 1-\tanh^2\frac{r-c_0}{c_2} \right]\;\;,
\end{eqnarray}
where  $c_0 = 0.491 23$, $c_1 =0.298 228$, $c_2 =0.198 739$, and $c_3 =0.521 298$. In the original benchmark study, this density profile is used as the EP density, while in our work we use it as the electron density. In gyrokinetic simulations, the temperature and density are given in terms of the Larmor radius and $\beta$ (the ratio of the plasma pressure to the magnetic pressure). While the nominal values are $\beta_e=8\pi n_eT_e/B_0^2=9\times10^{-4}$, $\rho_{N}=\SI{0.00152}{\meter}$ for the electromagnetic simulations, in this work, we adopt an enhanced density and temperature case with $\beta_e=0.03$, $\rho_N=7.8767\times10^{-3}\SI{}{\meter}$. Note that the value of $\beta_e$, however, does not change the values of the flux and bootstrap current normalized by the equilibrium density for the present model. 

\begin{figure}[h]
    \centering
    \includegraphics[width=.7\textwidth]{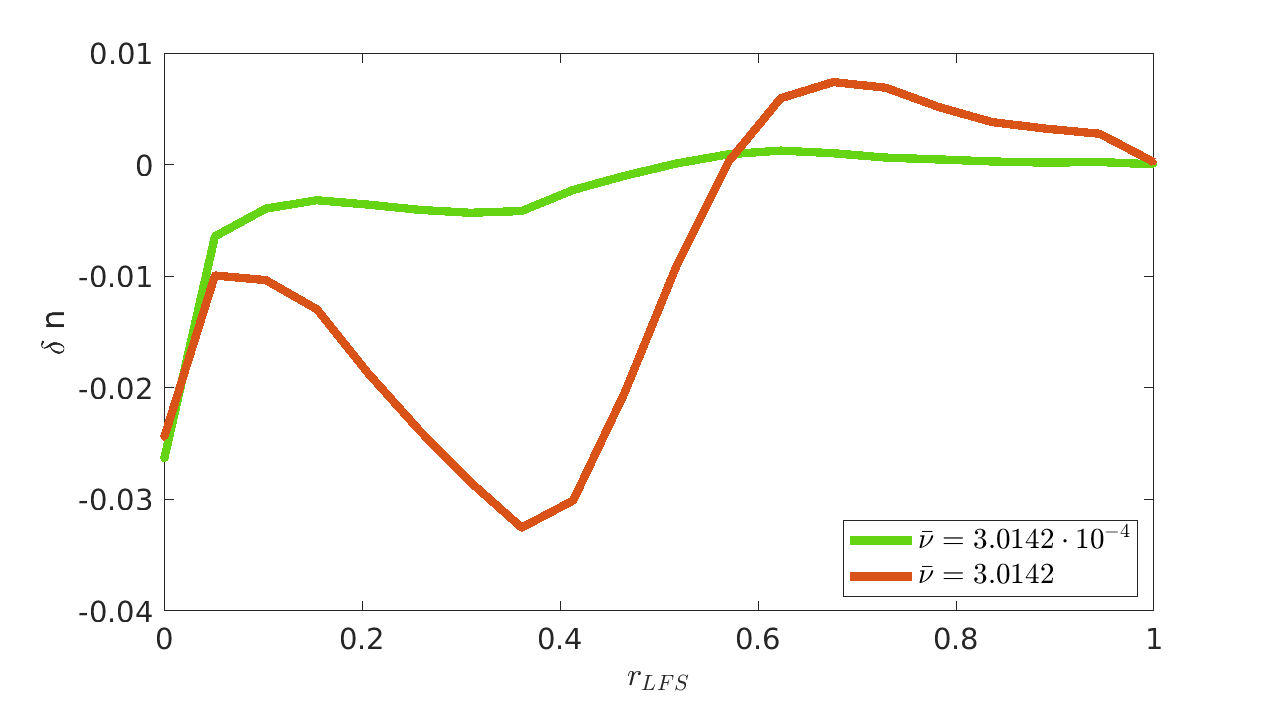}
    \caption{The radial profiles of the normalized density change at the end of the simulation, where $r_{LFS}$ is the radial coordinate from the axis to the direction of the low field side.}
    \label{fig:itpa-density-change}
\end{figure}

\begin{figure}[h]
    \centering
    \includegraphics[width=1\textwidth]{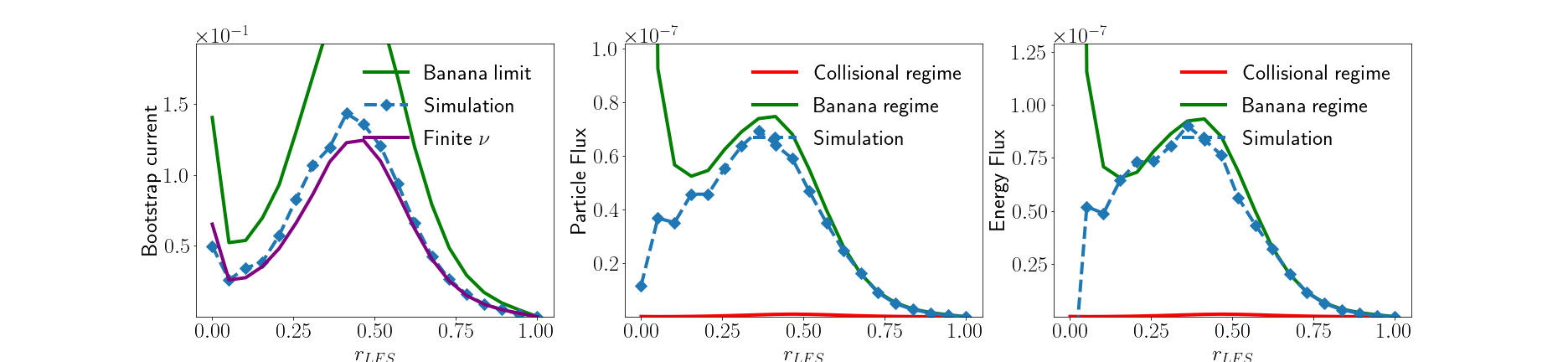}
    \caption{The radial profiles \lanaB{from the magnetic axis to the outer boundary of the simulation domain (at $r_{LFS}=\SI{1}{\meter}$) }of the bootstrap current, particle flux and energy flux for the low collision frequency case, where $\Bar{\nu}\approx3\cdot10^{-4}$. The dashed lines represent the simulation result, while the green and red lines represent the analytical solutions in the respective limits, while the \lanaB{purple} line is from the interpolation formula given in Eq. \ref{eq:finite-normalized-jb}. \zhixinC{Note that the interpolation formula for the bootstrap current (denoted as ``finite $\nu$'' in the figure) gives the current with the effect of finite collisionality in both low and high collisionalities. } The values of $\nu_*\approx0.05 \text{ to } 0.001$ correspond to radial locations of $r_{LFS}=0.1 \text{ to } \SI{1}{\meter}$, hence it is expected to be in the banana regime for all radial locations.}
    \label{fig:itpa-flux-jb-profiles-low-collision}
\end{figure}

\begin{figure}[h]
    \centering
    \includegraphics[width=1\textwidth]{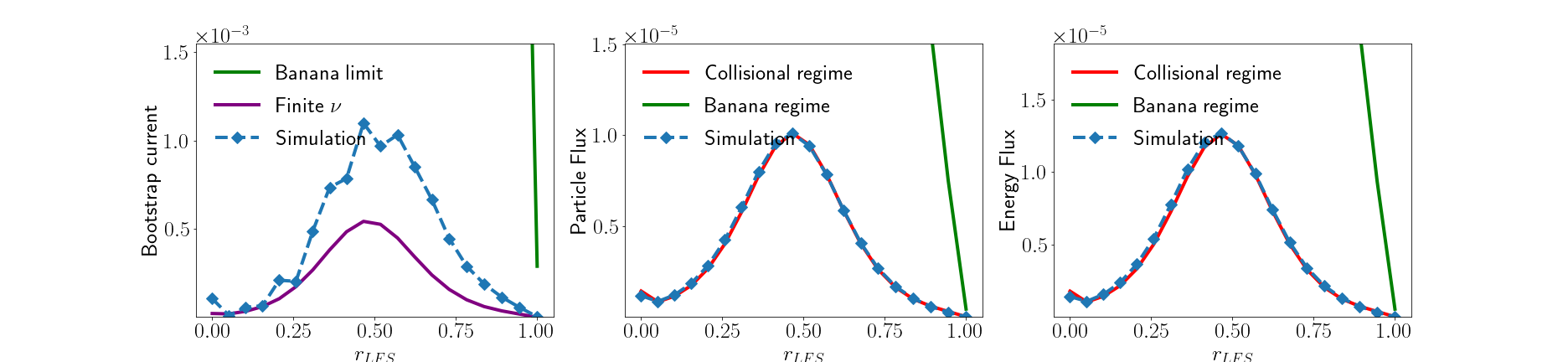}
    \caption{The radial profiles \lanaB{from the magnetic axis to the outer boundary of the simulation domain(at $r_{LFS}=\SI{1}{\meter}$) }of the bootstrap current, particle flux and energy flux for the high collision frequency case, where $\Bar{\nu}\approx3$. The dashed lines represent the simulation result, while the green and red lines represent the analytical solutions in the respective limits, while the \lanaB{purple} line is the interpolation formula given in Eq. \ref{eq:finite-normalized-jb}. The values of $\nu_*\approx\lanaB{500} \text{ to } 178$ correspond to radial locations of $r_{LFS}=0.1 \text{ to } \SI{1}{\meter}$, hence it is expected to be in the collisional regime for all radial locations.}
    \label{fig:itpa-flux-jb-profiles-high-collision}
\end{figure}

We simulate cases with high and low collision frequencies and compare the results with the local theory. \lanaB{The time step sizes were chosen such that $\nu\,\Delta t$ ranges from $6.7 \cdot 10^{-6}$ to $0.034$.} During the simulation, the density profile changes gradually because of the particle transport due to collisions. At the end of the simulation, the density profiles are analyzed, as can be seen in Fig. \ref{fig:itpa-density-change}. For the high collision case, the density change is of the order of $\sim3\%$, while for the low collision case, \lanaB{the change is $\sim2.5\%$ near the axis and $|\delta n|<1\%$ in the rest of the volume}. This density variation has negligible effects on the density gradient as given in Eqs. \ref{eq:itpa-density-profiles} with the chosen coefficients and thus the particle and energy fluxes and the bootstrap current stay at the same level after a ramp-up phase until the end of the simulations.
\lanaA{In addition, as our simulations are not in the long time scales we do not consider the known issues of unknown marker distribution in the Monte-Carlo delta-$f$ collision operators \cite{chen2022evolution} \cite{wang2004global}. However, for longer time scale simulations the deviation of the initial marker distribution for the initial one and the noise level also need to be systematically analysed in the future.}
The radial profiles of particle flux, energy flux, and bootstrap current are also analyzed, as shown in Figs. \ref{fig:itpa-flux-jb-profiles-low-collision}-\ref{fig:itpa-flux-jb-profiles-high-collision}. For the low collision frequency limit, we observe good agreement between the theoretical calculation and the simulation, as seen in Fig. \ref{fig:itpa-flux-jb-profiles-low-collision}. Blue dashed lines indicate the results from particle simulations. The green lines for the particle flux and energy flux indicate solutions in the low collisionality limit given by Eqs. \ref{eq:banana-normalized-Gamma}-\ref{eq:banana-normalized-Q}, while the red lines are in the high collisionality limit given by Eqs. \ref{eq:col-normalized-Gamma}-\ref{eq:col-normalized-Q}. 
\zhixinB{Note that although the value of the collision is set to be as close as possible to the collisionless or collisional limit in the simulations, it is still not the limit of $\nu\rightarrow0$ and $\nu\rightarrow\infty$, as adopted in the theoretical formulae, which can lead to minor or moderate discrepancies between the simulation and theoretical results, as shown in Figs.  \ref{fig:itpa-flux-jb-profiles-low-collision}-\ref{fig:itpa-flux-jb-profiles-high-collision} and other cases in the Cyclone case and the ASDEX-Upgrade case. }
As for the bootstrap current, the green line indicates the low collision limit given by Eq. \ref{eq:banana-normalized-jb}, and the orange line is given by the analytical formula by Hinton which takes into account finite collision frequency given in Eqs. \ref{eq:finite-normalized-jb} with coefficients in Eqs. \ref{coeff-Hinton1} - \ref{coeff-Hinton2}. 
We observe a big difference in the energy and particle flux near the axis for low collisionality. 
The discrepancy between the \zhixinA{standard neoclassical} theory and the simulation is also observed in the study of ion transport in previous work \cite{lin1997large}.
\zhixinB{Better agreement can be obtained in principle by considering the theoretical formula closer to the axis following the derivation in the neoclassical theory \cite{hinton1976theory,chang1982effect,sauter1999neoclassical} or by comparing with other codes \cite{wang2004global,lin1997large,belli2008kinetic} but it is beyond the scope of this work.
} 

\begin{figure}[h]
    \centering
    \includegraphics[width=1\textwidth]{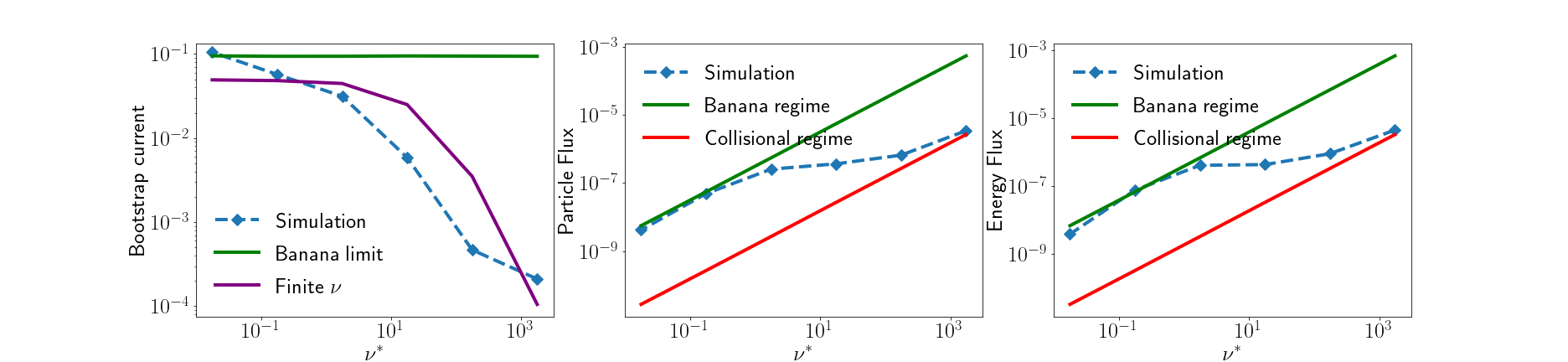}
    \caption{The maximum value of the bootstrap current, particle flux and energy flux as functions of collision frequency, taken at radial location $r=\SI{0.23}{\meter}$. \lanaB{For the bootstrap current, due to higher noise levels each value represents the average of three closest radial locations to $r=\SI{0.23}{\meter}$.} The blue dashed lines represent the simulation result, while the green and red lines represent the analytical solutions in the respective limits, while the \lanaB{purple} line is the interpolation formula given in Eq. \ref{eq:finite-normalized-jb}.}
    \label{fig:itpa-flux-jb-versus-collision}
\end{figure}

As for the high collisionality case, the results are given in Fig. \ref{fig:itpa-flux-jb-profiles-high-collision}. For the particle flux and the energy flux, a very good agreement with the analytical solutions is observed. 
The bootstrap current is much lower than that in the banana limit and thus more markers ($10^{7}$) are used in this simulation to enhance the signal-to-noise ratio.
The bootstap current from the simulation is of the same order of magnitude as but higher than the analytical result.
We also investigate the fluxes and bootstrap current as a function of the collision frequency as shown in Fig. \ref{fig:itpa-flux-jb-versus-collision}. The reference radial location was chosen arbitrarily to be  \AlanaB{$r\approx\SI{0.23}{\meter}$}, and the fluxes and the current are compared with the values from the theoretical radial profile at this reference location. 
We observe that the bootstrap current from simulations decreases much faster as collision frequency increases than expected from the analytical interpolation solution in the plateau and the collisional regimes. 
However, we observe good agreement for the particle and energy fluxes with the analytical solutions, in the low ($\nu^*<1$) and high ($\nu^*>10$) collisionality limits. Furthermore, the plateau ($1<\nu^*<10$) regime is also visible where the fluxes stay almost constant for different collision frequencies, and the overall behavior is qualitatively consistent with previous results  \cite{hinton1976theory,lin1996gyrokinetic}. 

\subsection{Electron transport results for the moderate aspect ratio case}
\label{sec:cyclone_results}
In this section, we study the electron transport for the moderate aspect ratio case featured by $a/R_0=0.36$. 
The details of the magnetic equilibrium and the density profile are listed in the reference \cite{lu2019development} and the main parameters are briefly summarized as follows. The major radius and the minor radius are $R_0=1.67$ m and $a=0.6012$ m, respectively. 
In generating the EQDSK file as the input of the magnetic equilibrium, the $\Bar{q}$ profile in Eq. \ref{eq:qbar1d} is adopted with 
\begin{eqnarray}
    \Bar{q}(r_c)=1.41\;\;,\;\;
    \frac{r}{\Bar{q}}\frac{d\Bar{q}}{dr} (r_c) = 0.8369 \;\;,
\end{eqnarray}
where $r_c=0.5a$. 
The radial profile of the density and its gradient are given analytically as follows \cite{lu2019development},
\begin{eqnarray}
\label{eq:nprofiles}
    n(r)&=&\exp\left\{-\kappa_n W_n\frac{a}{L_{ref}}\tanh\left(\frac{r-r_c}{W_na}\right)\right\}; \\
\label{eq:dlnndr_cyclone}
    \frac{d\ln n}{dr}&=&-\frac{1}{L_{ref}}\kappa_n\cosh^{-2}\left(\frac{r-r_c}{W_na}\right); 
\end{eqnarray}
where $L_{ref}=R_0$, $W_n=0.3$, $\kappa_n=2.23$.
$\beta_e=0.03$, $\rho_N=7.8767\times10^{-3}$ m.
For the sake of simplicity, a uniform temperature profile is assumed. 
Compared with the ITPA case ($R_0/a=10$, $q\approx1.75$ at $r/a=0.5$) studied in Chapter \ref{subsec:itpa_results}, the aspect ratio of the Cyclone case is smaller ($a/R_0=0.36$). In addition, the safety factor is lower than that of the ITPA case in the inner radial region ($r<0.5a$) but is larger near the edge. In the theoretical derivation, the small parameter $\rho_p=(qR_0/r)m_sv_\perp/(Z_seB)$ is used as the expansion parameter, where $Z_s$ is the charge number, and the subscript `s' indicates species `s'. As a result, for different values of $q$ and $r/R_0$, the accuracy of the theoretical result can be different, which is more relevant for ion transport. As $\rho_p$ is close to or even larger than the characteristic length of the equilibrium or the density/temperature profiles, the traditional neoclassical formulae are not valid and corrections are needed as shown in previous studies of ion transport \cite{chang1982effect,helander1998bifurcated}. For electron transport, $\rho_p\ll1$ is usually well satisfied except if it is very close to the magnetic axis where $R_0/r\rightarrow \infty$ and the traditional theoretical formulae can also break down. \lanaB{The time step sizes were chosen such that $\nu\,\Delta t$ ranges from $1.2 \cdot 10^{-4}$ to $0.12$, which is the highest value among the three cases, hence in the highest collisionality case non-physical effects from the implementation of the collision operator need to be considered. The verification of the convergence of this $\nu\,\Delta t$ value is shown on the right-hand side of Fig.\ref{fig:convergence}.}

\begin{figure}[h]
    \centering
    \includegraphics[width=.7\textwidth]{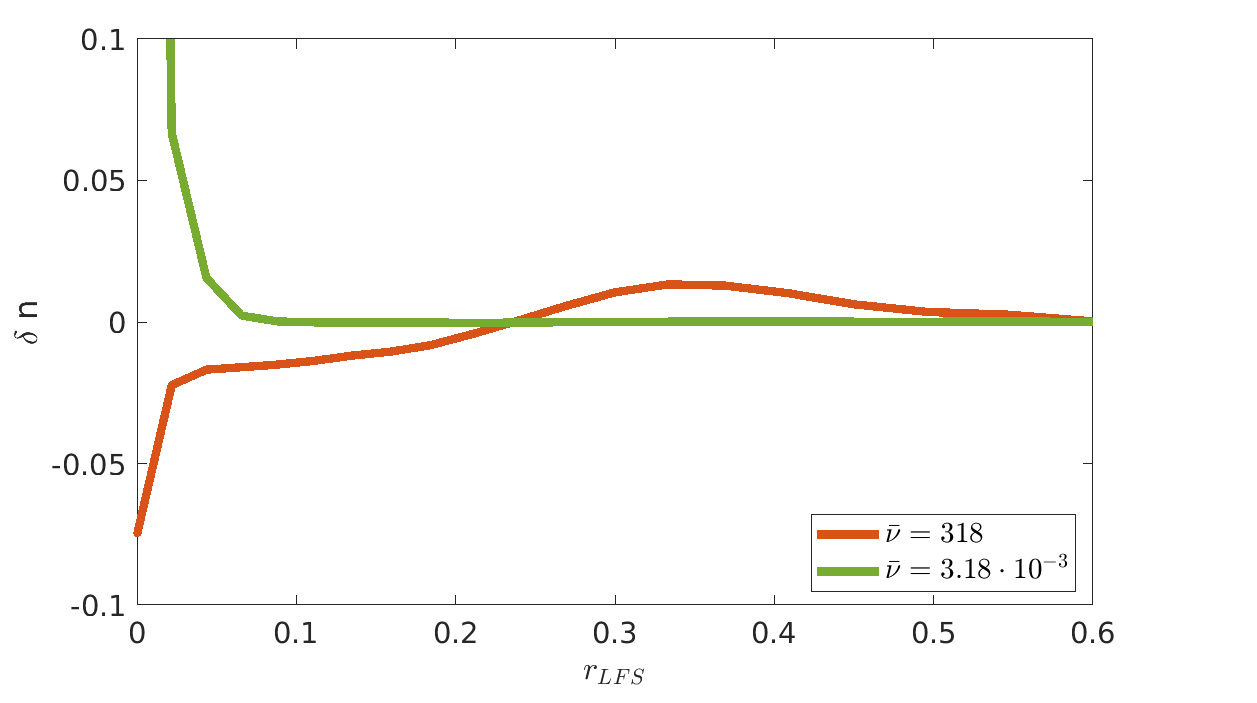}
    \caption{The radial profiles of the normalized density change at the end of the simulation, where $r_{LFS}$ indicates the radial coordinate from the axis to the low field side.}
    \label{fig:cyclone-density-change}
\end{figure}

In our simulation, we observe changes in the density of the order of $0.01\%$ in the low collision frequency case, and about $2\%$ in the high collisional case, \zhixinC{as shown in Fig. \ref{fig:cyclone-density-change}. } 
We also observe large density changes on the axis, which can be due to the discontinuity of the density profile near the axis (the radial gradient of the density profile is not exactly zero at the axis according to Eq. \ref{eq:dlnndr_cyclone} and can cause non-physical $\partial\ln n/\partial R$ and $\partial\ln n/\partial Z$ values in the 2D interpolation). \zhixinC{The positive $\delta n$ near the axis for the low collisionality can be due to the different parameter conditions near the axis such as $r/R_0$ and the consequent physical collisionality. } Nevertheless, the on-axis physics is not the focus of this work and it does not affect the physics near the middle radius due to the small orbit width of electrons.

\begin{figure}[h]
    \centering
    \includegraphics[width=1\textwidth]{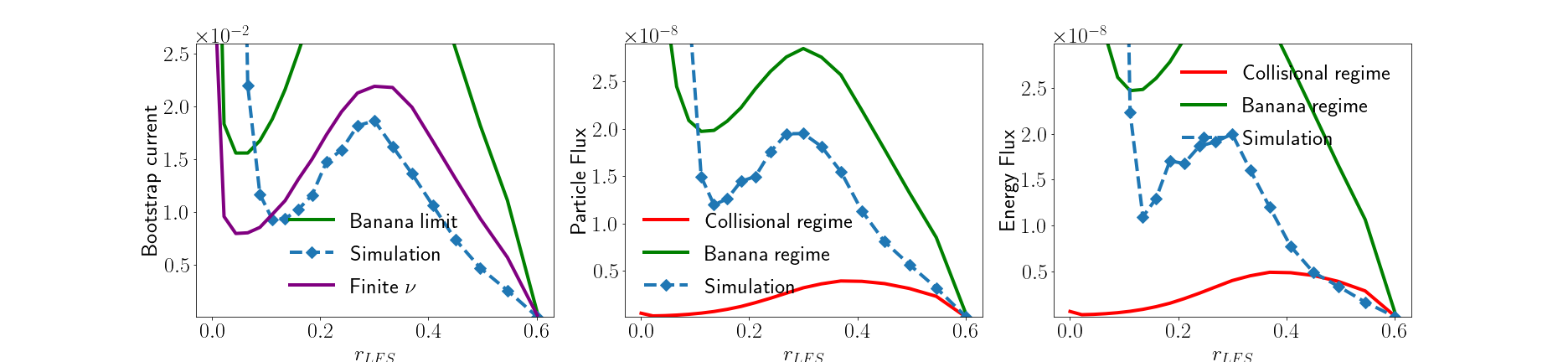}
    \caption{The radial profiles \lanaB{from the magnetic axis to the outer boundary of the simulation domain (at $r_{LFS}=\SI{0.6}{\meter}$) }of the bootstrap current, particle flux and energy flux for the low collision frequency case, where $\Bar{\nu}\approx3\cdot10^{-3}$. The blue dashed lines represent the simulation result, while the green and red lines represent the analytical solutions in the respective limits and the \lanaB{purple} line is the interpolation formula given in Eq. \ref{eq:finite-normalized-jb}. The values of $\nu_*\approx0.04  \text{ to } 0.004$ correspond to radial locations of $r_{LFS}=0.1 \text{ to } \SI{0.6}{\meter}$, hence it is expected to be in the banana regime for all radial locations.}
    \label{fig:cyclone-flux-jb-profiles-low-collision}
\end{figure}

\begin{figure}[h]
    \centering
    \includegraphics[width=1\textwidth]{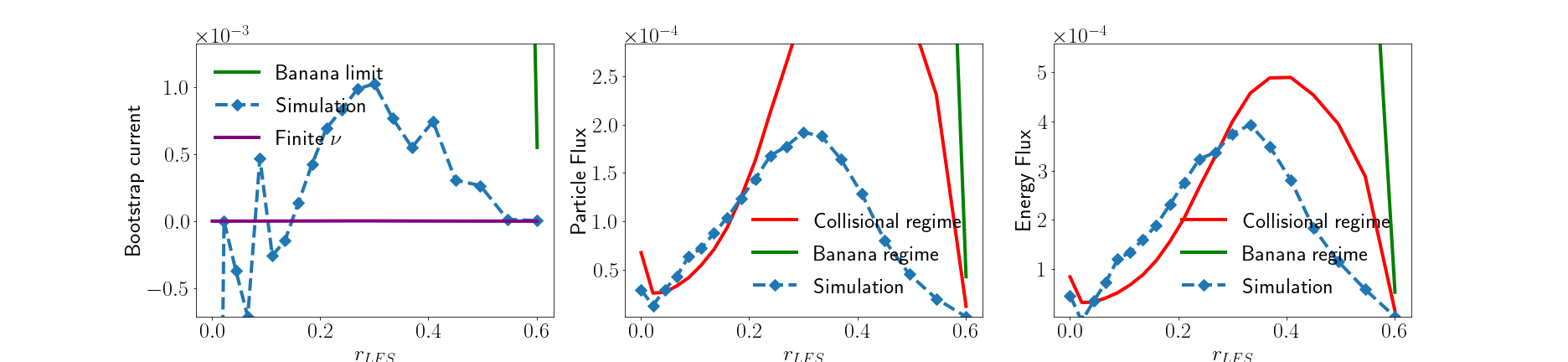}
    \caption{The radial profiles \lanaB{from the magnetic axis to the outer boundary of the simulation domain (at $r_{LFS}=\SI{0.6}{\meter}$) }of the bootstrap current, particle flux and energy flux for the high collision frequency case, where $\Bar{\nu}\approx3\cdot10^{2}$. The dashed blue lines represent the simulation result, while the green and red lines represent the analytical solutions in the respective limits and the \lanaB{purple} line is the interpolation formula given in Eq. \ref{eq:finite-normalized-jb}. The values of $\nu_*\approx4800  \text{ to } 400$ correspond to radial locations of $r_{LFS}=0.1 \text{ to } \SI{0.6}{\meter}$, hence it is expected to be in the collisional regime for all radial locations.}
    \label{fig:cyclone-flux-jb-profiles-high-collision}
\end{figure}

We again start by analyzing the radial particle flux, energy flux, and bootstrap current profiles of high and low collision frequency cases, as shown in Figs. \ref{fig:cyclone-flux-jb-profiles-low-collision}--\ref{fig:cyclone-flux-jb-profiles-high-collision}. For the low collisionality case, the agreement is not as good as that in the ITPA large aspect ratio case. The reason can be the large aspect ratio approximation adopted in the analytical formulas and the different treatments in our code for the poloidal-angle dependent values such as $q$ in calculating the theoretical fluxes and the current. \lanaB{It merits more effort in the future to adopt analytical formulas for comparision which are valid for broad regimes of aspect ratios\cite{chang1986effect} and for general axisymmetric equilibria and all collisionality regimes \cite{sauter1999neoclassical}. }
The local theory for the comparison in this work is derived in the large aspect ratio limit and \zhixinB{the good agreement is observed in this parameter regime in the ITPA case.}
In addition, when calculating the theoretical values of the fluxes and current, we calculate the theoretical values on numerous points in one annulus; then the averaged value is calculated with proper weights and thus the flux surface average value is obtained using the Monte-Carlo integration. This method of calculating the flux-surface-averaged fluxes and current gives us a convenient and practical way of calculating the fluxes and current for shaped tokamak geometry, as we also adopted for the ASDEX Upgrade case in the next section.
For the high collision frequency case, the agreement is better closer to the axis, which is expected as the local aspect ratio is larger. 
The bootstrap current is much larger than the analytical solution. The reason can be the approximation in the interpolation formula Eq. \ref{eq:finite-normalized-jb}, which is derived to match the results at the low and high collision regime. Indeed, the discrepancies between the transition formula and the formulae at the low and high collisions are also observed theoretically \cite{hinton1976theory}.
In the theoretical solutions, the interpolation formula is obtained by fitting the analytical results in the banana-plateau and plateau-collisional regimes, hence our simulation result is expected to be more exact than the analytical solution close to the plateau regime. However, there are analytical theories 

\begin{figure}[h]
    \centering
    \includegraphics[width=1\textwidth]{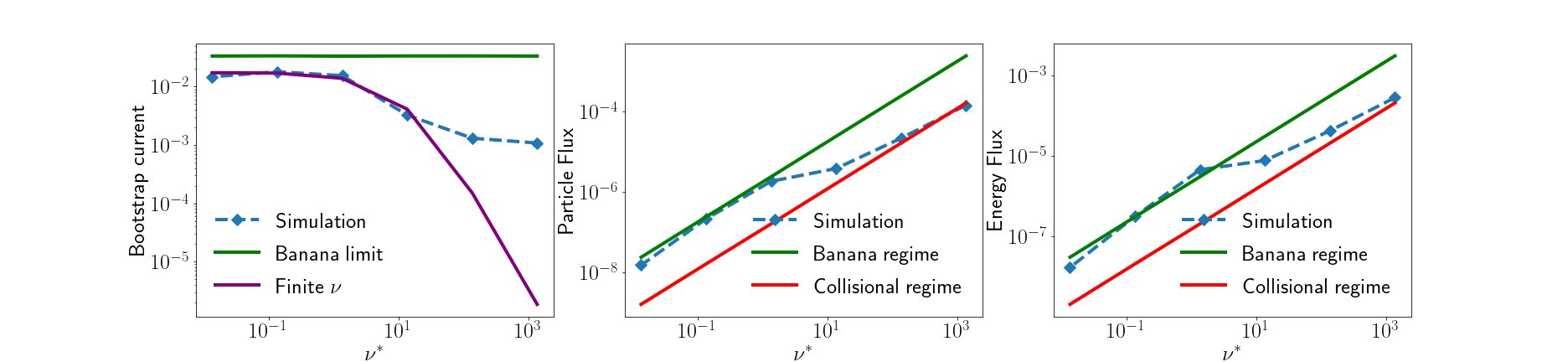}
    \caption{The values of the bootstrap current, particle flux and energy flux for $r_{LFS}=\SI{0.21}{\meter}$ and different values of collision frequency. The blue lines represent the simulation result, while the green and red lines represent the analytical solutions in the respective limits and the \lanaB{purple} line is the interpolation formula given in Eq. \ref{eq:finite-normalized-jb}. \lanaB{As the choice of the radial location is arbitrary, this agreement only compares this specific radial location to the theory. For radial locations closer to the axis the agreement would be better, while further away from the middle radius ($r_{LFS}=\SI{0.3}{\meter}$) the agreement would be worse.}}
    \label{fig:cyclone-flux-jb-versus-collision}
\end{figure}

Analyzing how fluxes change with different collision frequencies, we take the radial location close to the maximum value of the fluxes ($r=\SI{0.21}{\meter}$), which in this case is also a larger aspect ratio case. We observe good agreement with the analytical results for the particle and energy fluxes, as shown in Fig. \ref{fig:cyclone-flux-jb-versus-collision}. The bootstrap current from the simulation follows the interpolation formula (red line) well. For $\nu^*>10^2$, the discrepancy between the simulation and the interpolation formula is larger and it can be due to the inaccuracy of the interpolation formula. 
For the bootstrap current, the lowest collision frequency case does not produce the highest current as would be expected. This is due to the bootstrap current still increasing slightly when the simulation is finished. We keep the lowest collision case as the indicator of the most expensive case that we can study with our present capability. Note that in this study of the dependency on collisionality, the collision frequency varies by a factor of $10^5$ and the needed minimum simulation time also varies significantly since several ($\sim10$) collisional periods are needed to reach the saturated state of the fluxes and the current. The lowest collisional case is the most expensive one in order to observe reasonable results.
The current simulation consumes 14 hours of computation time on 4 nodes with each node containing two Intel(R) Xeon(R) Gold 6130 processors (16 cores per processor, 2.10GHz, 22MB Cache). 
Both the simulation results and the theoretical interpolation formula results are lower than the collisionless approximation and have the expected behavior of decreasing when the collision frequency is increasing.

\subsection{Electron transport results for the ASDEX Upgrade case}
\label{subsec:aug_results}

\begin{figure}[h]
    \centering
    \includegraphics[width=0.8\textwidth]{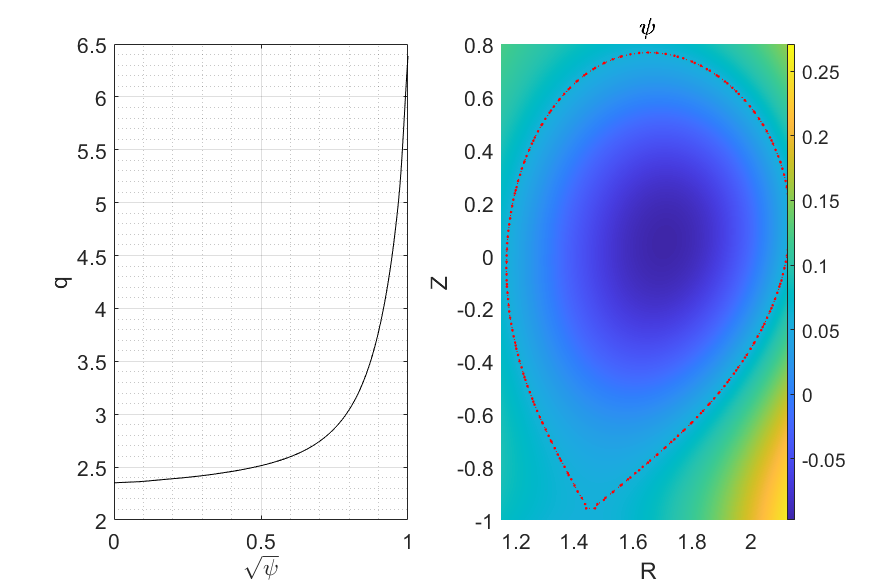}
    \caption{The q profiles (left) and poloidal magnetic flux map (right) of the AUG case. \lanaB{The dashed red line represents the last closed magnetic surface.}}
    \label{fig:aug-q-profiles}
\end{figure}

In this section, a realistic geometry for Tokamak plasmas is used. The
ASDEX Upgrade (AUG) case with shot number 34924 at {3.600} s is
chosen as adopted by the previous work for the development of the TRIMEG code for the studies of the ion temperature gradient mode \cite{lu2019development}. This is a typical discharge for the study of energetic particles
and turbulence physics \cite{lauber2018strongly}. The EQDSK file is obtained from experimental data. \lanaB{The major radius $R_0\approx\SI{1.71}{\meter}$. }The $q$ profile and the poloidal magnetic flux function are shown in Fig. \ref{fig:aug-q-profiles}.
In the simulation, we use the experimental equilibrium but uniform temperature and the analytical density profiles in Eq. \ref{eq:nprofiles} with the radial coordinate replaced with
$\rho_{pol}=\sqrt{(\psi-\psi_0)/(\psi_b-\psi_0)}$, where the subscript $0$ and $b$ indicate the values at the magnetic axis and the last closed surface, respectively. 
$\beta_e=0.03$, $\rho_N=\SI{0.01}{\meter}$.
In this work, we focus on testing the capability of the code in treating realistic geometry with minimum technical complexity.
\lanaB{The time step sizes were chosen such that $\nu\,\Delta t$ ranges from $2.78 \cdot 10^{-5}$ to $0.0278$.}

\begin{figure}[h]
    \centering
    \includegraphics[width=.7\textwidth]{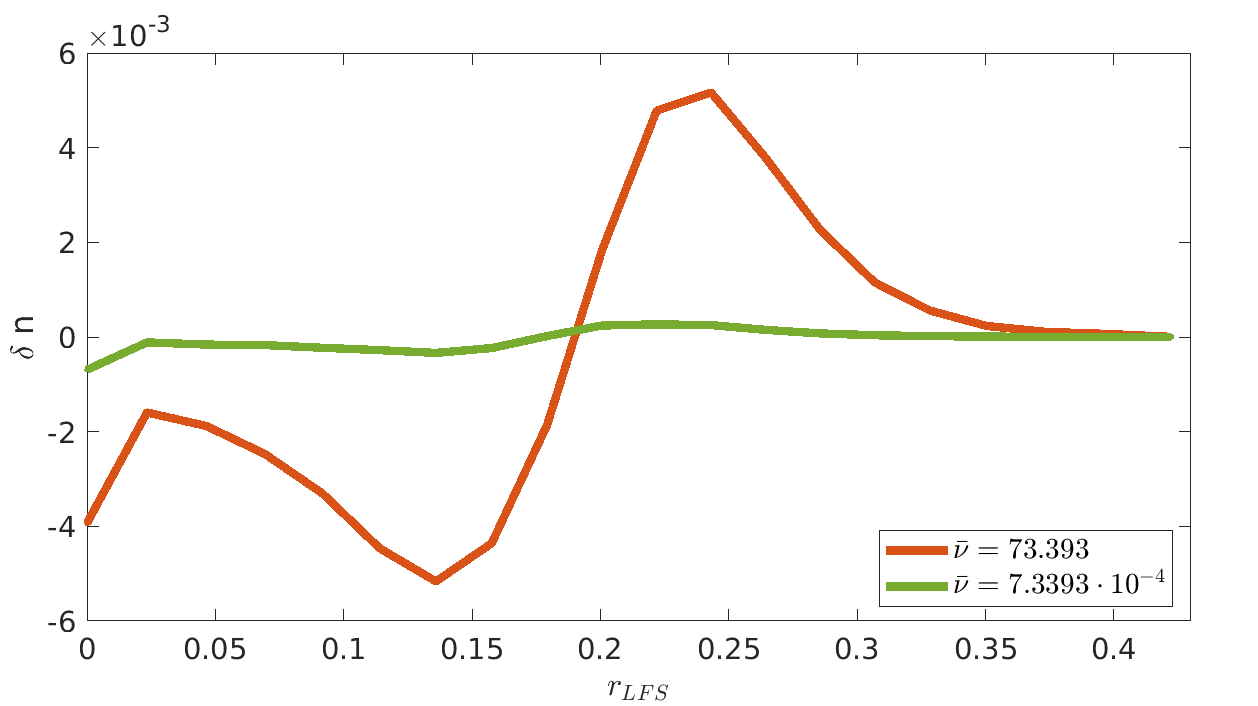}
    \caption{The radial profiles of the normalized density change at the end of the simulation, where $r_{LFS}$ indicates the radial coordinate from the axis to the low field side.}
    \label{fig:aug-density-change}
\end{figure}

As in previous chapters, we first look at the density changes due to low and high collision frequencies as shown in Fig. \ref{fig:aug-density-change}. The density changes are much smaller than in the previous cases. The density change due to high collision frequency is $0.6\%$, while in the low collision frequency case, it is almost negligible. 

\begin{figure}[h]
    \centering
    \includegraphics[width=1\textwidth]{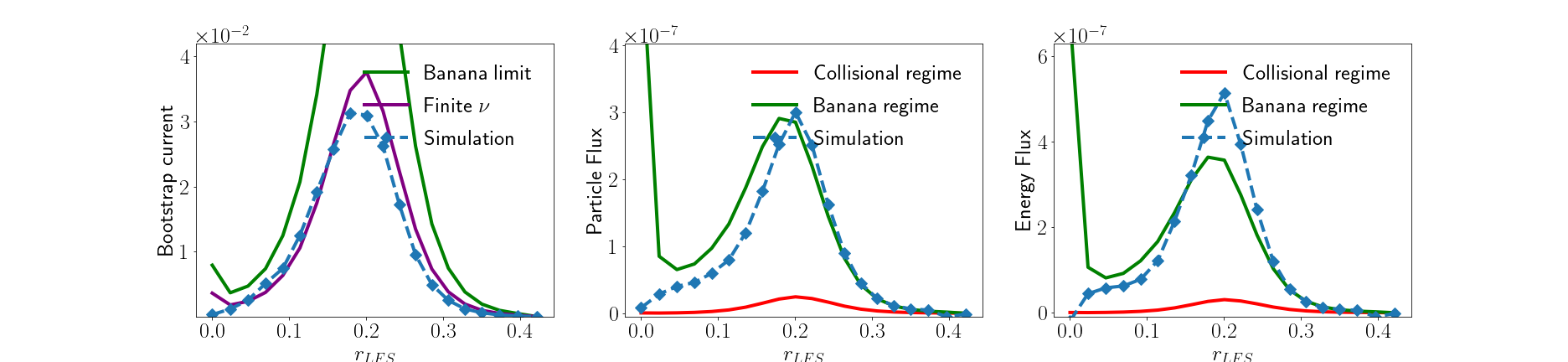}
    \caption{The radial profiles \lanaB{from the magnetic axis to the outer boundary of the simulation domain (at $r_{LFS}=\SI{0.4}{\meter}$) }of the bootstrap current, particle flux and energy flux for the low collision case, where \zhixinC{the collision frequency (normalized to the code unit) is} $\Bar{\nu}\approx7\cdot10^{-3}$. The dashed blue lines represent the simulation result, while the green and red lines represent the analytical solutions in the respective limits and the \lanaB{purple} line is the interpolation formula given in Eq. \ref{eq:finite-normalized-jb}. \zhixinC{Corresponding to the value of $\bar\nu$,} the values of \AlanaB{$\nu_*\approx0.2  \text{ to } 0.03$} correspond to radial locations of $r_{LFS}=0.1 \text{ to } \SI{0.4}{\meter}$, hence it is expected to be in the banana regime for all radial locations. \zhixinC{Note that $\bar{\nu}$ is given but not calculated based on the plasma parameters for the purpose of benchmark.}}
    \label{fig:aug-flux-jb-profiles-low-collision}
\end{figure}

\begin{figure}[h]
    \centering
    \includegraphics[width=1\textwidth]{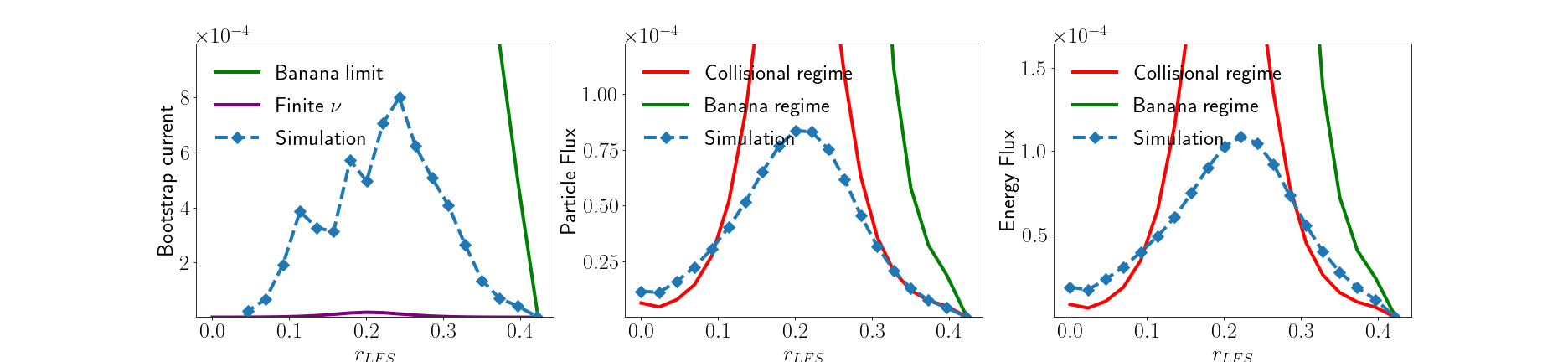}
    \caption{The radial profiles \lanaB{from the magnetic axis to the outer boundary of the simulation domain (at $r_{LFS}=\SI{0.4}{\meter}$) }of the bootstrap current, particle flux and energy flux for the high collision case, where $\Bar{\nu}\approx7\cdot10^{1}$. The dashed blue lines represent the simulation result, while the green and red lines represent the analytical solutions in the respective limits and the \lanaB{purple} line is the interpolation formula given in Eq. \ref{eq:finite-normalized-jb}. The values of \AlanaB{$\nu_*\approx2120  \text{ to } 265$} correspond to radial locations of $r_{LFS}=0.1 \text{ to } \SI{0.4}{\meter}$, hence it is expected to be in the collisional regime for all radial locations.}
    \label{fig:aug-flux-jb-profiles-high-collision}
\end{figure}

Looking at the radial profile for the low collision case given in Fig. \ref{fig:aug-flux-jb-profiles-low-collision}, we observe good agreement with the theory. The bootstrap current and the particle flux agree very well, while the energy flux has a higher value at the center than predicted by the neoclassical theory.
As for the high collision frequency case, given in Fig. \ref{fig:aug-flux-jb-profiles-high-collision}, the discrepancy between theory and simulation is larger than that in the ITPA and Cyclone cases. In addition to the reason we discussed in the Cyclone case, for the AUG case, the magnetic flux surfaces are not circular but are strongly shaped. The theoretical formulae of fluxes and bootstrap current were derived for the circular magnetic flux surface originally \cite{hinton1976theory} and our scheme of flux surface average of the fluxes and current is one possible way of an estimate, which is more reasonable for circular magnetic flux surfaces and is for the verification of the implementation of the code. More accurate theoretical/numerical solutions for shaped tokamak plasmas can be found elsewhere \cite{sauter1999neoclassical,belli2008kinetic} and the comparison with our simulation results is possible, but is beyond the scope of this work. 

\begin{figure}[h]
    \centering
    \includegraphics[width=1\textwidth]{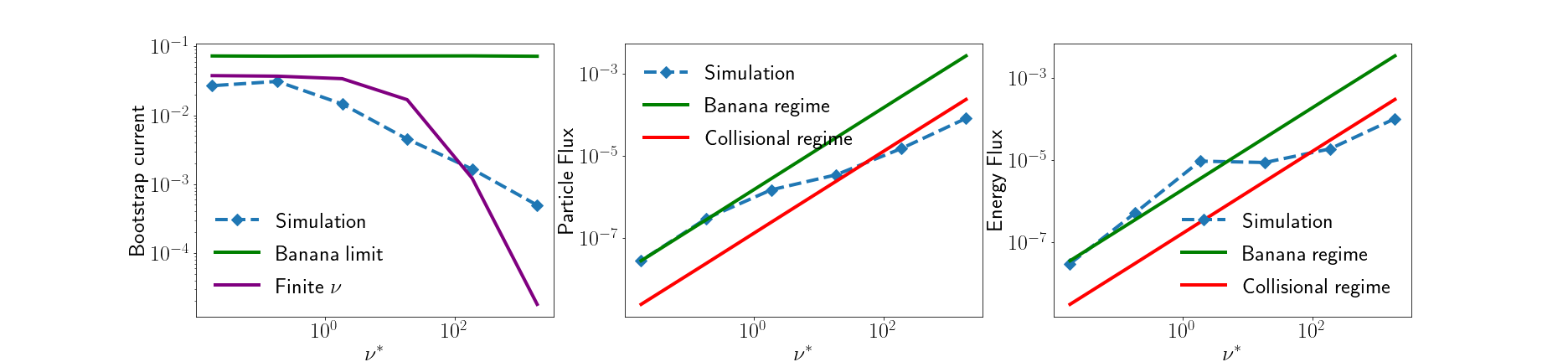}
    \caption{The values of the bootstrap current, particle flux and energy flux for different values of collision frequency at the low field side radius $r_{LFS}=\SI{0.2}{\meter}$. The dashed blue lines represent the simulation result, while the green and red lines represent the analytical solutions in the respective limits and the \lanaB{purple} line is the interpolation formula given in Eq. \ref{eq:finite-normalized-jb}. \lanaB{As the choice of the radial location is arbitrary, this agreement only compares this specific radial location to the theory. For radial locations closer to the middle radius ($r_{LFS}=\SI{0.2}{\meter}$) the agreement would be worse, while near to the axis and further away from the middle radius the agreement is better.}}
    \label{fig:aug-flux-jb-versus-collision}
\end{figure}

Furthermore, to investigate the dependence on the collision frequency we picked the radial coordinate where the highest values of fluxes were observed ($r_{LFS}=\SI{0.2}{\meter}$), and compared the analytical and simulation results, as shown in Fig. \ref{fig:aug-flux-jb-versus-collision}. The agreement with the theory can still be seen. However, the agreement for large collision frequencies is worse. Nevertheless, the trends of the fluxes and bootstrap current follow the theoretical results. More issues can be studied in the future for understanding the connections and the differences between the global gyro-kinetic simulation and the local theory, in order to identify the origin of the differences between the theoretical results and the simulation results as well as the limitation of the local theory. \lanaA{For future studies, comparing results to other gyrokinetic codes, such as ORB5 \cite{lanti2019global} or GENE \cite{Gorler2012GENE} would be the next step for verification in specific cases.} \lanaC{Aditionally, actual experimental electron tranport levels in the ASDEX Upgrade can also be compared \cite{ryter2003electron,meyer2019overview}, however this merits more effort as contributions from both turbulence and neoclassical physics need to be considered.}

\section{Conclusion and outlook}
\label{sec:conclusions}
We studied the electron transport and bootstrap current generation by adding a pitch angle scattering operator to the TRIMEG code, which uses an unstructured mesh and equations in $(R,Z)$ coordinates. This provides a robust tool in a broad collisionality range and for flexible parameters such as tokamak geometry. In this work, we only considered the electron species, set the temperature gradient to zero, but take into account the density gradients. We first compared the simulation results to analytical calculations in the large aspect ratio ($R_0/a=10$) approximation and found good agreement in those limits between theory and simulation. For moderate aspect ratio ($R_0/a=1/0.36$) cases the agreement was worse, as expected, due to the approximations in the theory such as the large aspect ratio approximation.
In the case with AUG experimental geometry, in the collisional regime, the radial profiles of bootstrap current and fluxes are simulated, with significant discrepancy between the local theory and the simulation results. 
In the banana regime, the agreement between the theory and the simulation results are reasonably good.
In the studies of collisionality scan, for high collision frequencies, the energy and particle fluxes decrease in magnitude faster than the analytical solutions as the collision frequency increased.

This work has verified the capability of the TRIMEG code to study the electron transport and the bootstrap current generation in tokamak plasmas for further simulations with open field lines.
Future steps for further investigation would be to investigate the neoclassical physics in realistic experimental geometry with experimental density and temperature profiles. It also merits more efforts to add like-particle collision operators \AlanaB{ \cite{lin1995gyrokinetic,wang2004global} }\lanaB{ and fully non-linear collision operators \cite{hager2016gyrokinetic} for more realistic studies}. The particle-field coupling also needs to be added, for the study of the neoclassical radial electric field in future, which plays a key role in the instability stabilization and the optimization of the confinement performance.
\begin{acknowledgments}
This work has been carried out within the framework of the EUROfusion Consortium, funded by the European Union via the Euratom Research and Training Programme (Grant Agreement No 101052200 -- EUROfusion). Views and opinions expressed are however those of the author(s) only and do not necessarily reflect those of the European Union or the European Commission. Neither the European Union nor the European Commission can be held responsible for them.
\end{acknowledgments}
\zhixinB{
\appendix
\section{Results related to numerical convergences}
The main concern in the numerical convergence is related to the marker number and the time step size. We select two typical cases to demonstrate the principles we adopted when choosing values of the time step size and the marker number. 
While it is not practical to perform convergence studies for each case, the bootstrap current, particle flux and energy flux are compared for different values of the relevant parameters.
As shown in the left frame of Fig. \ref{fig:convergence}, we calculate the particle flux at the middle minor radius by interpolation ($r/a=0.5$). 
For the ITPA case \AlanaB{(}$\bar\nu\approx 3$, $\nu\Delta t=0.00581$), results are compared for different marker numbers as shown in the left frame. The results start to converge as $N_p\ge 10^6$.
As showin in the right frame of Fig. \ref{fig:convergence}, the particle fluxes are compared with different time step sizes. For this ASDEX Upgrade case ($\bar\nu\approx7\cdot10^1$), the results start to converge as $\nu\Delta t\le0.11133$. In our simulations, we choose $\nu\Delta t\le0.055664$ for $\bar{\nu}\approx7\cdot10^1$ and smaller $dt$ for other cases of ASDEX Upgrade. 
\begin{figure}[h]
    \centering
    \includegraphics[width=.33\textwidth]{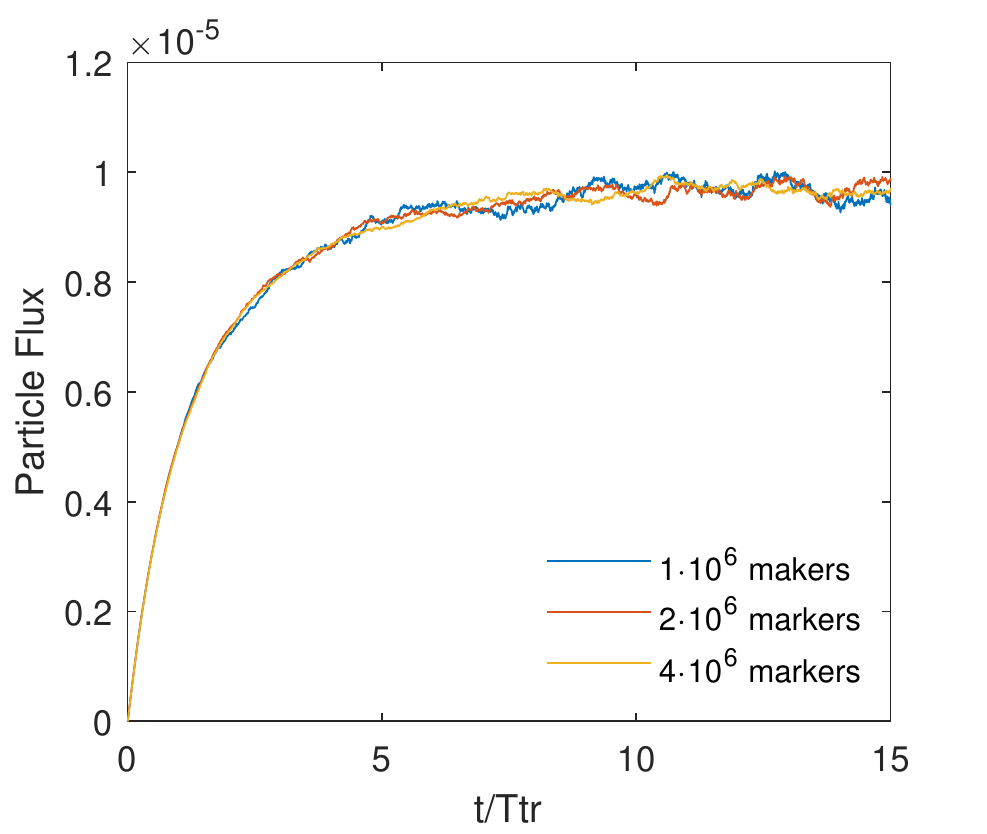}
    \includegraphics[width=.33\textwidth]{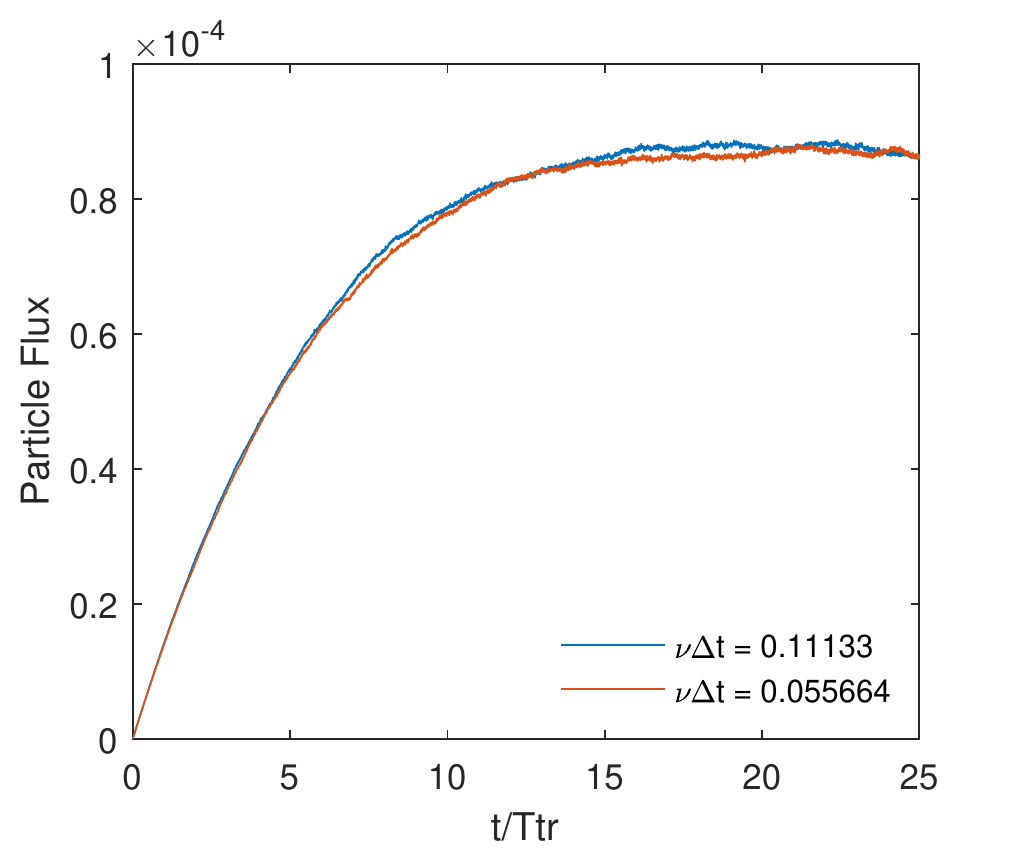}
    \caption{The time evolution of the particle flux at $r/a=0.5$ for the ITPA-TAE case (left) and the ASDEX-Upgrade case (right).}
    \label{fig:convergence}
\end{figure}
}
\bibliography{references}

\begin{thebibliography}{31}%
\makeatletter
\providecommand \@ifxundefined [1]{%
 \@ifx{#1\undefined}
}%
\providecommand \@ifnum [1]{%
 \ifnum #1\expandafter \@firstoftwo
 \else \expandafter \@secondoftwo
 \fi
}%
\providecommand \@ifx [1]{%
 \ifx #1\expandafter \@firstoftwo
 \else \expandafter \@secondoftwo
 \fi
}%
\providecommand \natexlab [1]{#1}%
\providecommand \enquote  [1]{``#1''}%
\providecommand \bibnamefont  [1]{#1}%
\providecommand \bibfnamefont [1]{#1}%
\providecommand \citenamefont [1]{#1}%
\providecommand \href@noop [0]{\@secondoftwo}%
\providecommand \href [0]{\begingroup \@sanitize@url \@href}%
\providecommand \@href[1]{\@@startlink{#1}\@@href}%
\providecommand \@@href[1]{\endgroup#1\@@endlink}%
\providecommand \@sanitize@url [0]{\catcode `\\12\catcode `\$12\catcode
  `\&12\catcode `\#12\catcode `\^12\catcode `\_12\catcode `\%12\relax}%
\providecommand \@@startlink[1]{}%
\providecommand \@@endlink[0]{}%
\providecommand \url  [0]{\begingroup\@sanitize@url \@url }%
\providecommand \@url [1]{\endgroup\@href {#1}{\urlprefix }}%
\providecommand \urlprefix  [0]{URL }%
\providecommand \Eprint [0]{\href }%
\providecommand \doibase [0]{http://dx.doi.org/}%
\providecommand \selectlanguage [0]{\@gobble}%
\providecommand \bibinfo  [0]{\@secondoftwo}%
\providecommand \bibfield  [0]{\@secondoftwo}%
\providecommand \translation [1]{[#1]}%
\providecommand \BibitemOpen [0]{}%
\providecommand \bibitemStop [0]{}%
\providecommand \bibitemNoStop [0]{.\EOS\space}%
\providecommand \EOS [0]{\spacefactor3000\relax}%
\providecommand \BibitemShut  [1]{\csname bibitem#1\endcsname}%
\let\auto@bib@innerbib\@empty
\bibitem [{\citenamefont {Beidler}\ \emph {et~al.}(2021)\citenamefont
  {Beidler}, \citenamefont {Smith}, \citenamefont {Alonso}, \citenamefont
  {Andreeva}, \citenamefont {Baldzuhn}, \citenamefont {Beurskens},
  \citenamefont {Borchardt}, \citenamefont {Bozhenkov}, \citenamefont
  {Brunner}, \citenamefont {Damm} \emph {et~al.}}]{beidler2021demonstration}%
  \BibitemOpen
  \bibfield  {author} {\bibinfo {author} {\bibfnamefont {C.}~\bibnamefont
  {Beidler}}, \bibinfo {author} {\bibfnamefont {H.}~\bibnamefont {Smith}},
  \bibinfo {author} {\bibfnamefont {A.}~\bibnamefont {Alonso}}, \bibinfo
  {author} {\bibfnamefont {T.}~\bibnamefont {Andreeva}}, \bibinfo {author}
  {\bibfnamefont {J.}~\bibnamefont {Baldzuhn}}, \bibinfo {author}
  {\bibfnamefont {M.}~\bibnamefont {Beurskens}}, \bibinfo {author}
  {\bibfnamefont {M.}~\bibnamefont {Borchardt}}, \bibinfo {author}
  {\bibfnamefont {S.}~\bibnamefont {Bozhenkov}}, \bibinfo {author}
  {\bibfnamefont {K.}~\bibnamefont {Brunner}}, \bibinfo {author} {\bibfnamefont
  {H.}~\bibnamefont {Damm}},  \emph {et~al.},\ }\href@noop {} {\bibfield
  {journal} {\bibinfo  {journal} {Nature}\ }\textbf {\bibinfo {volume} {596}},\
  \bibinfo {pages} {221} (\bibinfo {year} {2021})}\BibitemShut {NoStop}%
\bibitem [{\citenamefont {Helander}(1998)}]{helander1998bifurcated}%
  \BibitemOpen
  \bibfield  {author} {\bibinfo {author} {\bibfnamefont {P.}~\bibnamefont
  {Helander}},\ }\href@noop {} {\bibfield  {journal} {\bibinfo  {journal}
  {Physics of Plasmas}\ }\textbf {\bibinfo {volume} {5}},\ \bibinfo {pages}
  {3999} (\bibinfo {year} {1998})}\BibitemShut {NoStop}%
\bibitem [{\citenamefont {Hinton}\ and\ \citenamefont
  {Hazeltine}(1976)}]{hinton1976theory}%
  \BibitemOpen
  \bibfield  {author} {\bibinfo {author} {\bibfnamefont {F.}~\bibnamefont
  {Hinton}}\ and\ \bibinfo {author} {\bibfnamefont {R.}~\bibnamefont
  {Hazeltine}},\ }\href@noop {} {\bibfield  {journal} {\bibinfo  {journal}
  {Reviews of Modern Physics}\ }\textbf {\bibinfo {volume} {48}},\ \bibinfo
  {pages} {239} (\bibinfo {year} {1976})}\BibitemShut {NoStop}%
\bibitem [{\citenamefont {Hirshman}\ and\ \citenamefont
  {Sigmar}(1981)}]{hirshman1981neoclassical}%
  \BibitemOpen
  \bibfield  {author} {\bibinfo {author} {\bibfnamefont {S.}~\bibnamefont
  {Hirshman}}\ and\ \bibinfo {author} {\bibfnamefont {D.}~\bibnamefont
  {Sigmar}},\ }\href@noop {} {\bibfield  {journal} {\bibinfo  {journal}
  {Nuclear Fusion}\ }\textbf {\bibinfo {volume} {21}},\ \bibinfo {pages} {1079}
  (\bibinfo {year} {1981})}\BibitemShut {NoStop}%
\bibitem [{\citenamefont {Lin}, \citenamefont {Tang},\ and\ \citenamefont
  {Lee}(1997)}]{lin1997large}%
  \BibitemOpen
  \bibfield  {author} {\bibinfo {author} {\bibfnamefont {Z.}~\bibnamefont
  {Lin}}, \bibinfo {author} {\bibfnamefont {W.}~\bibnamefont {Tang}}, \ and\
  \bibinfo {author} {\bibfnamefont {W.}~\bibnamefont {Lee}},\ }\href@noop {}
  {\bibfield  {journal} {\bibinfo  {journal} {Physics of Plasmas}\ }\textbf
  {\bibinfo {volume} {4}},\ \bibinfo {pages} {1707} (\bibinfo {year}
  {1997})}\BibitemShut {NoStop}%
\bibitem [{\citenamefont {Wang}\ \emph
  {et~al.}(2006{\natexlab{a}})\citenamefont {Wang}, \citenamefont {Rewoldt},
  \citenamefont {Tang}, \citenamefont {Hinton}, \citenamefont {Manickam},
  \citenamefont {Zakharov}, \citenamefont {White},\ and\ \citenamefont
  {Kaye}}]{wang2006nonlocal}%
  \BibitemOpen
  \bibfield  {author} {\bibinfo {author} {\bibfnamefont {W.}~\bibnamefont
  {Wang}}, \bibinfo {author} {\bibfnamefont {G.}~\bibnamefont {Rewoldt}},
  \bibinfo {author} {\bibfnamefont {W.}~\bibnamefont {Tang}}, \bibinfo {author}
  {\bibfnamefont {F.}~\bibnamefont {Hinton}}, \bibinfo {author} {\bibfnamefont
  {J.}~\bibnamefont {Manickam}}, \bibinfo {author} {\bibfnamefont
  {L.}~\bibnamefont {Zakharov}}, \bibinfo {author} {\bibfnamefont
  {R.}~\bibnamefont {White}}, \ and\ \bibinfo {author} {\bibfnamefont
  {S.}~\bibnamefont {Kaye}},\ }\href@noop {} {\bibfield  {journal} {\bibinfo
  {journal} {Physics of plasmas}\ }\textbf {\bibinfo {volume} {13}},\ \bibinfo
  {pages} {082501} (\bibinfo {year} {2006}{\natexlab{a}})}\BibitemShut
  {NoStop}%
\bibitem [{\citenamefont {Chang}, \citenamefont {Ku},\ and\ \citenamefont
  {Weitzner}(2004)}]{chang2004numerical}%
  \BibitemOpen
  \bibfield  {author} {\bibinfo {author} {\bibfnamefont {C.}~\bibnamefont
  {Chang}}, \bibinfo {author} {\bibfnamefont {S.}~\bibnamefont {Ku}}, \ and\
  \bibinfo {author} {\bibfnamefont {H.}~\bibnamefont {Weitzner}},\ }\href@noop
  {} {\bibfield  {journal} {\bibinfo  {journal} {Physics of Plasmas}\ }\textbf
  {\bibinfo {volume} {11}},\ \bibinfo {pages} {2649} (\bibinfo {year}
  {2004})}\BibitemShut {NoStop}%
\bibitem [{\citenamefont {Zhao}, \citenamefont {Chankin},\ and\ \citenamefont
  {Coster}(2019)}]{zhao2019solps}%
  \BibitemOpen
  \bibfield  {author} {\bibinfo {author} {\bibfnamefont {M.}~\bibnamefont
  {Zhao}}, \bibinfo {author} {\bibfnamefont {A.}~\bibnamefont {Chankin}}, \
  and\ \bibinfo {author} {\bibfnamefont {D.}~\bibnamefont {Coster}},\
  }\href@noop {} {\bibfield  {journal} {\bibinfo  {journal} {Plasma Physics and
  Controlled Fusion}\ }\textbf {\bibinfo {volume} {61}},\ \bibinfo {pages}
  {025019} (\bibinfo {year} {2019})}\BibitemShut {NoStop}%
\bibitem [{\citenamefont {Lee}(1983)}]{lee1983gyrokinetic}%
  \BibitemOpen
  \bibfield  {author} {\bibinfo {author} {\bibfnamefont {W.}~\bibnamefont
  {Lee}},\ }\href@noop {} {\bibfield  {journal} {\bibinfo  {journal} {The
  Physics of Fluids}\ }\textbf {\bibinfo {volume} {26}},\ \bibinfo {pages}
  {556} (\bibinfo {year} {1983})}\BibitemShut {NoStop}%
\bibitem [{\citenamefont {Lin}, \citenamefont {Tang},\ and\ \citenamefont
  {Lee}(1995)}]{lin1995gyrokinetic}%
  \BibitemOpen
  \bibfield  {author} {\bibinfo {author} {\bibfnamefont {Z.}~\bibnamefont
  {Lin}}, \bibinfo {author} {\bibfnamefont {W.}~\bibnamefont {Tang}}, \ and\
  \bibinfo {author} {\bibfnamefont {W.}~\bibnamefont {Lee}},\ }\href@noop {}
  {\bibfield  {journal} {\bibinfo  {journal} {Physics of Plasmas}\ }\textbf
  {\bibinfo {volume} {2}},\ \bibinfo {pages} {2975} (\bibinfo {year}
  {1995})}\BibitemShut {NoStop}%
\bibitem [{\citenamefont {Bergmann}, \citenamefont {Peeters},\ and\
  \citenamefont {Pinches}(2001)}]{bergmann2001guiding}%
  \BibitemOpen
  \bibfield  {author} {\bibinfo {author} {\bibfnamefont {A.}~\bibnamefont
  {Bergmann}}, \bibinfo {author} {\bibfnamefont {A.}~\bibnamefont {Peeters}}, \
  and\ \bibinfo {author} {\bibfnamefont {S.}~\bibnamefont {Pinches}},\
  }\href@noop {} {\bibfield  {journal} {\bibinfo  {journal} {Physics of
  Plasmas}\ }\textbf {\bibinfo {volume} {8}},\ \bibinfo {pages} {5192}
  (\bibinfo {year} {2001})}\BibitemShut {NoStop}%
\bibitem [{\citenamefont {Wang}\ \emph
  {et~al.}(2006{\natexlab{b}})\citenamefont {Wang}, \citenamefont {Lin},
  \citenamefont {Tang}, \citenamefont {Lee}, \citenamefont {Ethier},
  \citenamefont {Lewandowski}, \citenamefont {Rewoldt}, \citenamefont {Hahm},\
  and\ \citenamefont {Manickam}}]{wang2006gyro}%
  \BibitemOpen
  \bibfield  {author} {\bibinfo {author} {\bibfnamefont {W.}~\bibnamefont
  {Wang}}, \bibinfo {author} {\bibfnamefont {Z.}~\bibnamefont {Lin}}, \bibinfo
  {author} {\bibfnamefont {W.}~\bibnamefont {Tang}}, \bibinfo {author}
  {\bibfnamefont {W.}~\bibnamefont {Lee}}, \bibinfo {author} {\bibfnamefont
  {S.}~\bibnamefont {Ethier}}, \bibinfo {author} {\bibfnamefont
  {J.}~\bibnamefont {Lewandowski}}, \bibinfo {author} {\bibfnamefont
  {G.}~\bibnamefont {Rewoldt}}, \bibinfo {author} {\bibfnamefont
  {T.}~\bibnamefont {Hahm}}, \ and\ \bibinfo {author} {\bibfnamefont
  {J.}~\bibnamefont {Manickam}},\ }\href@noop {} {\bibfield  {journal}
  {\bibinfo  {journal} {Physics of Plasmas}\ }\textbf {\bibinfo {volume}
  {13}},\ \bibinfo {pages} {092505} (\bibinfo {year}
  {2006}{\natexlab{b}})}\BibitemShut {NoStop}%
\bibitem [{\citenamefont {Vernay}\ \emph {et~al.}(2010)\citenamefont {Vernay},
  \citenamefont {Brunner}, \citenamefont {Villard}, \citenamefont {McMillan},
  \citenamefont {Jolliet}, \citenamefont {Tran}, \citenamefont {Bottino},\ and\
  \citenamefont {Graves}}]{vernay2010neoclassical}%
  \BibitemOpen
  \bibfield  {author} {\bibinfo {author} {\bibfnamefont {T.}~\bibnamefont
  {Vernay}}, \bibinfo {author} {\bibfnamefont {S.}~\bibnamefont {Brunner}},
  \bibinfo {author} {\bibfnamefont {L.}~\bibnamefont {Villard}}, \bibinfo
  {author} {\bibfnamefont {B.}~\bibnamefont {McMillan}}, \bibinfo {author}
  {\bibfnamefont {S.}~\bibnamefont {Jolliet}}, \bibinfo {author} {\bibfnamefont
  {T.}~\bibnamefont {Tran}}, \bibinfo {author} {\bibfnamefont {A.}~\bibnamefont
  {Bottino}}, \ and\ \bibinfo {author} {\bibfnamefont {J.}~\bibnamefont
  {Graves}},\ }\href@noop {} {\bibfield  {journal} {\bibinfo  {journal}
  {Physics of Plasmas}\ }\textbf {\bibinfo {volume} {17}},\ \bibinfo {pages}
  {122301} (\bibinfo {year} {2010})}\BibitemShut {NoStop}%
\bibitem [{\citenamefont {Lu}\ \emph {et~al.}(2019)\citenamefont {Lu},
  \citenamefont {Lauber}, \citenamefont {Hayward-Schneider}, \citenamefont
  {Bottino},\ and\ \citenamefont {Hoelzl}}]{lu2019development}%
  \BibitemOpen
  \bibfield  {author} {\bibinfo {author} {\bibfnamefont {Z.}~\bibnamefont
  {Lu}}, \bibinfo {author} {\bibfnamefont {P.}~\bibnamefont {Lauber}}, \bibinfo
  {author} {\bibfnamefont {T.}~\bibnamefont {Hayward-Schneider}}, \bibinfo
  {author} {\bibfnamefont {A.}~\bibnamefont {Bottino}}, \ and\ \bibinfo
  {author} {\bibfnamefont {M.}~\bibnamefont {Hoelzl}},\ }\href@noop {}
  {\bibfield  {journal} {\bibinfo  {journal} {Physics of Plasmas}\ }\textbf
  {\bibinfo {volume} {26}},\ \bibinfo {pages} {122503} (\bibinfo {year}
  {2019})}\BibitemShut {NoStop}%
\bibitem [{\citenamefont {Lu}\ \emph {et~al.}(2021)\citenamefont {Lu},
  \citenamefont {Meng}, \citenamefont {Hoelzl},\ and\ \citenamefont
  {Lauber}}]{lu2021development}%
  \BibitemOpen
  \bibfield  {author} {\bibinfo {author} {\bibfnamefont {Z.}~\bibnamefont
  {Lu}}, \bibinfo {author} {\bibfnamefont {G.}~\bibnamefont {Meng}}, \bibinfo
  {author} {\bibfnamefont {M.}~\bibnamefont {Hoelzl}}, \ and\ \bibinfo {author}
  {\bibfnamefont {P.}~\bibnamefont {Lauber}},\ }\href@noop {} {\bibfield
  {journal} {\bibinfo  {journal} {Journal of Computational Physics}\ }\textbf
  {\bibinfo {volume} {440}},\ \bibinfo {pages} {110384} (\bibinfo {year}
  {2021})}\BibitemShut {NoStop}%
\bibitem [{\citenamefont {Lu}\ \emph {et~al.}(2023)\citenamefont {Lu},
  \citenamefont {Meng}, \citenamefont {Hatzky}, \citenamefont {Hoelzl},\ and\
  \citenamefont {Lauber}}]{Lu_2023}%
  \BibitemOpen
  \bibfield  {author} {\bibinfo {author} {\bibfnamefont {Z.}~\bibnamefont
  {Lu}}, \bibinfo {author} {\bibfnamefont {G.}~\bibnamefont {Meng}}, \bibinfo
  {author} {\bibfnamefont {R.}~\bibnamefont {Hatzky}}, \bibinfo {author}
  {\bibfnamefont {M.}~\bibnamefont {Hoelzl}}, \ and\ \bibinfo {author}
  {\bibfnamefont {P.}~\bibnamefont {Lauber}},\ }\href {\doibase
  10.1088/1361-6587/acb010} {\bibfield  {journal} {\bibinfo  {journal} {Plasma
  Physics and Controlled Fusion}\ }\textbf {\bibinfo {volume} {65}},\ \bibinfo
  {pages} {034004} (\bibinfo {year} {2023})}\BibitemShut {NoStop}%
\bibitem [{\citenamefont {Lin}(1996)}]{lin1996gyrokinetic}%
  \BibitemOpen
  \bibfield  {author} {\bibinfo {author} {\bibfnamefont {Z.}~\bibnamefont
  {Lin}},\ }\href@noop {} {\emph {\bibinfo {title} {Gyrokinetic particle
  simulations of neoclassical transport}}}\ (\bibinfo  {publisher} {Princeton
  University},\ \bibinfo {year} {1996})\BibitemShut {NoStop}%
\bibitem [{\citenamefont {Lanti}(2019)}]{lanti2019global}%
  \BibitemOpen
  \bibfield  {author} {\bibinfo {author} {\bibfnamefont {E.}~\bibnamefont
  {Lanti}},\ }\href@noop {} {\enquote {\bibinfo {title} {Global flux-driven
  simulations of ion temperature-gradient and trapped-electron modes driven
  turbulence with an improved multithreaded gyrokinetic pic code},}\ }\bibinfo
  {type} {Tech. Rep.}\ (\bibinfo  {institution} {EPFL},\ \bibinfo {year}
  {2019})\BibitemShut {NoStop}%
\bibitem [{\citenamefont {Hager}\ and\ \citenamefont
  {Chang}(2016)}]{hager2016gyrokinetic}%
  \BibitemOpen
  \bibfield  {author} {\bibinfo {author} {\bibfnamefont {R.}~\bibnamefont
  {Hager}}\ and\ \bibinfo {author} {\bibfnamefont {C.}~\bibnamefont {Chang}},\
  }\href@noop {} {\bibfield  {journal} {\bibinfo  {journal} {Phys. Plasmas}\
  }\textbf {\bibinfo {volume} {23}},\ \bibinfo {pages} {042503} (\bibinfo
  {year} {2016})}\BibitemShut {NoStop}%
\bibitem [{\citenamefont {Hazeltine}\ and\ \citenamefont
  {Waelbroeck}(2019)}]{hazeltine2019framework}%
  \BibitemOpen
  \bibfield  {author} {\bibinfo {author} {\bibfnamefont {R.}~\bibnamefont
  {Hazeltine}}\ and\ \bibinfo {author} {\bibfnamefont {F.}~\bibnamefont
  {Waelbroeck}},\ }\href@noop {} {\emph {\bibinfo {title} {The Framework of
  Plasma Physics}}},\ Frontiers in Physics\ (\bibinfo  {publisher} {Taylor \&
  Francis Group},\ \bibinfo {year} {2019})\BibitemShut {NoStop}%
\bibitem [{\citenamefont {K{\"o}nies}\ \emph {et~al.}(2018)\citenamefont
  {K{\"o}nies}, \citenamefont {Briguglio}, \citenamefont {Gorelenkov},
  \citenamefont {Feh{\'e}r}, \citenamefont {Isaev}, \citenamefont {Lauber},
  \citenamefont {Mishchenko}, \citenamefont {Spong}, \citenamefont {Todo},
  \citenamefont {Cooper} \emph {et~al.}}]{konies2018benchmark}%
  \BibitemOpen
  \bibfield  {author} {\bibinfo {author} {\bibfnamefont {A.}~\bibnamefont
  {K{\"o}nies}}, \bibinfo {author} {\bibfnamefont {S.}~\bibnamefont
  {Briguglio}}, \bibinfo {author} {\bibfnamefont {N.}~\bibnamefont
  {Gorelenkov}}, \bibinfo {author} {\bibfnamefont {T.}~\bibnamefont
  {Feh{\'e}r}}, \bibinfo {author} {\bibfnamefont {M.}~\bibnamefont {Isaev}},
  \bibinfo {author} {\bibfnamefont {P.}~\bibnamefont {Lauber}}, \bibinfo
  {author} {\bibfnamefont {A.}~\bibnamefont {Mishchenko}}, \bibinfo {author}
  {\bibfnamefont {D.~A.}\ \bibnamefont {Spong}}, \bibinfo {author}
  {\bibfnamefont {Y.}~\bibnamefont {Todo}}, \bibinfo {author} {\bibfnamefont
  {W.~A.}\ \bibnamefont {Cooper}},  \emph {et~al.},\ }\href@noop {} {\bibfield
  {journal} {\bibinfo  {journal} {Nuclear Fusion}\ }\textbf {\bibinfo {volume}
  {58}},\ \bibinfo {pages} {126027} (\bibinfo {year} {2018})}\BibitemShut
  {NoStop}%
\bibitem [{\citenamefont {Chen}, \citenamefont {Cheng},\ and\ \citenamefont
  {Parker}(2022)}]{chen2022evolution}%
  \BibitemOpen
  \bibfield  {author} {\bibinfo {author} {\bibfnamefont {Y.}~\bibnamefont
  {Chen}}, \bibinfo {author} {\bibfnamefont {J.}~\bibnamefont {Cheng}}, \ and\
  \bibinfo {author} {\bibfnamefont {S.~E.}\ \bibnamefont {Parker}},\
  }\href@noop {} {\bibfield  {journal} {\bibinfo  {journal} {Physics of
  Plasmas}\ }\textbf {\bibinfo {volume} {29}},\ \bibinfo {pages} {073901}
  (\bibinfo {year} {2022})}\BibitemShut {NoStop}%
\bibitem [{\citenamefont {Wang}\ \emph {et~al.}(2004)\citenamefont {Wang},
  \citenamefont {Tang}, \citenamefont {Hinton}, \citenamefont {Zakharov},
  \citenamefont {White},\ and\ \citenamefont {Manickam}}]{wang2004global}%
  \BibitemOpen
  \bibfield  {author} {\bibinfo {author} {\bibfnamefont {W.}~\bibnamefont
  {Wang}}, \bibinfo {author} {\bibfnamefont {W.}~\bibnamefont {Tang}}, \bibinfo
  {author} {\bibfnamefont {F.}~\bibnamefont {Hinton}}, \bibinfo {author}
  {\bibfnamefont {L.~E.}\ \bibnamefont {Zakharov}}, \bibinfo {author}
  {\bibfnamefont {R.}~\bibnamefont {White}}, \ and\ \bibinfo {author}
  {\bibfnamefont {J.}~\bibnamefont {Manickam}},\ }\href@noop {} {\bibfield
  {journal} {\bibinfo  {journal} {Computer physics communications}\ }\textbf
  {\bibinfo {volume} {164}},\ \bibinfo {pages} {178} (\bibinfo {year}
  {2004})}\BibitemShut {NoStop}%
\bibitem [{\citenamefont {Chang}\ and\ \citenamefont
  {Hinton}(1982)}]{chang1982effect}%
  \BibitemOpen
  \bibfield  {author} {\bibinfo {author} {\bibfnamefont {C.}~\bibnamefont
  {Chang}}\ and\ \bibinfo {author} {\bibfnamefont {F.}~\bibnamefont {Hinton}},\
  }\href@noop {} {\bibfield  {journal} {\bibinfo  {journal} {The Physics of
  Fluids}\ }\textbf {\bibinfo {volume} {25}},\ \bibinfo {pages} {1493}
  (\bibinfo {year} {1982})}\BibitemShut {NoStop}%
\bibitem [{\citenamefont {Sauter}, \citenamefont {Angioni},\ and\ \citenamefont
  {Lin-Liu}(1999)}]{sauter1999neoclassical}%
  \BibitemOpen
  \bibfield  {author} {\bibinfo {author} {\bibfnamefont {O.}~\bibnamefont
  {Sauter}}, \bibinfo {author} {\bibfnamefont {C.}~\bibnamefont {Angioni}}, \
  and\ \bibinfo {author} {\bibfnamefont {Y.}~\bibnamefont {Lin-Liu}},\
  }\href@noop {} {\bibfield  {journal} {\bibinfo  {journal} {Physics of
  Plasmas}\ }\textbf {\bibinfo {volume} {6}},\ \bibinfo {pages} {2834}
  (\bibinfo {year} {1999})}\BibitemShut {NoStop}%
\bibitem [{\citenamefont {Belli}\ and\ \citenamefont
  {Candy}(2008)}]{belli2008kinetic}%
  \BibitemOpen
  \bibfield  {author} {\bibinfo {author} {\bibfnamefont {E.}~\bibnamefont
  {Belli}}\ and\ \bibinfo {author} {\bibfnamefont {J.}~\bibnamefont {Candy}},\
  }\href@noop {} {\bibfield  {journal} {\bibinfo  {journal} {Plasma Physics and
  Controlled Fusion}\ }\textbf {\bibinfo {volume} {50}},\ \bibinfo {pages}
  {095010} (\bibinfo {year} {2008})}\BibitemShut {NoStop}%
\bibitem [{\citenamefont {Chang}\ and\ \citenamefont
  {Hinton}(1986)}]{chang1986effect}%
  \BibitemOpen
  \bibfield  {author} {\bibinfo {author} {\bibfnamefont {C.-S.}\ \bibnamefont
  {Chang}}\ and\ \bibinfo {author} {\bibfnamefont {F.}~\bibnamefont {Hinton}},\
  }\href@noop {} {\bibfield  {journal} {\bibinfo  {journal} {The Physics of
  fluids}\ }\textbf {\bibinfo {volume} {29}},\ \bibinfo {pages} {3314}
  (\bibinfo {year} {1986})}\BibitemShut {NoStop}%
\bibitem [{\citenamefont {Lauber}\ \emph {et~al.}(2018)\citenamefont {Lauber},
  \citenamefont {Geiger}, \citenamefont {Papp}, \citenamefont {Por},
  \citenamefont {Guimarais}, \citenamefont {Poloskei}, \citenamefont
  {Igochine}, \citenamefont {Maraschek}, \citenamefont {Pokol}, \citenamefont
  {Hayward-Schneider} \emph {et~al.}}]{lauber2018strongly}%
  \BibitemOpen
  \bibfield  {author} {\bibinfo {author} {\bibfnamefont {P.}~\bibnamefont
  {Lauber}}, \bibinfo {author} {\bibfnamefont {B.}~\bibnamefont {Geiger}},
  \bibinfo {author} {\bibfnamefont {G.}~\bibnamefont {Papp}}, \bibinfo {author}
  {\bibfnamefont {G.}~\bibnamefont {Por}}, \bibinfo {author} {\bibfnamefont
  {L.}~\bibnamefont {Guimarais}}, \bibinfo {author} {\bibfnamefont {P.~Z.}\
  \bibnamefont {Poloskei}}, \bibinfo {author} {\bibfnamefont {V.}~\bibnamefont
  {Igochine}}, \bibinfo {author} {\bibfnamefont {M.}~\bibnamefont {Maraschek}},
  \bibinfo {author} {\bibfnamefont {G.}~\bibnamefont {Pokol}}, \bibinfo
  {author} {\bibfnamefont {T.}~\bibnamefont {Hayward-Schneider}},  \emph
  {et~al.},\ }\href@noop {} {\bibfield  {journal} {\bibinfo  {journal}
  {proceedings of the 27th IAEA Fusion energy}\ } (\bibinfo {year}
  {2018})}\BibitemShut {NoStop}%
\bibitem [{\citenamefont {G{\"o}rler}\ \emph {et~al.}(2012)\citenamefont
  {G{\"o}rler}, \citenamefont {Lapillonne}, \citenamefont {Brunner},
  \citenamefont {Dannert}, \citenamefont {Jenko}, \citenamefont {Merz},\ and\
  \citenamefont {Told}}]{Gorler2012GENE}%
  \BibitemOpen
  \bibfield  {author} {\bibinfo {author} {\bibfnamefont {T.}~\bibnamefont
  {G{\"o}rler}}, \bibinfo {author} {\bibfnamefont {X.}~\bibnamefont
  {Lapillonne}}, \bibinfo {author} {\bibfnamefont {S.}~\bibnamefont {Brunner}},
  \bibinfo {author} {\bibfnamefont {T.}~\bibnamefont {Dannert}}, \bibinfo
  {author} {\bibfnamefont {F.}~\bibnamefont {Jenko}}, \bibinfo {author}
  {\bibfnamefont {F.}~\bibnamefont {Merz}}, \ and\ \bibinfo {author}
  {\bibfnamefont {D.}~\bibnamefont {Told}},\ }\href {\doibase
  10.1016/j.jcp.2011.05.034} {\bibfield  {journal} {\bibinfo  {journal}
  {Journal of Computational Physics}\ }\textbf {\bibinfo {volume} {230}},\
  \bibinfo {pages} {7053} (\bibinfo {year} {2012})}\BibitemShut {NoStop}%
\bibitem [{\citenamefont {Ryter}\ \emph {et~al.}(2003)\citenamefont {Ryter},
  \citenamefont {Tardini}, \citenamefont {De~Luca}, \citenamefont {Fahrbach},
  \citenamefont {Imbeaux}, \citenamefont {Jacchia}, \citenamefont {Kirov},
  \citenamefont {Leuterer}, \citenamefont {Mantica}, \citenamefont {Peeters}
  \emph {et~al.}}]{ryter2003electron}%
  \BibitemOpen
  \bibfield  {author} {\bibinfo {author} {\bibfnamefont {F.}~\bibnamefont
  {Ryter}}, \bibinfo {author} {\bibfnamefont {G.}~\bibnamefont {Tardini}},
  \bibinfo {author} {\bibfnamefont {F.}~\bibnamefont {De~Luca}}, \bibinfo
  {author} {\bibfnamefont {H.-U.}\ \bibnamefont {Fahrbach}}, \bibinfo {author}
  {\bibfnamefont {F.}~\bibnamefont {Imbeaux}}, \bibinfo {author} {\bibfnamefont
  {A.}~\bibnamefont {Jacchia}}, \bibinfo {author} {\bibfnamefont
  {K.}~\bibnamefont {Kirov}}, \bibinfo {author} {\bibfnamefont
  {F.}~\bibnamefont {Leuterer}}, \bibinfo {author} {\bibfnamefont
  {P.}~\bibnamefont {Mantica}}, \bibinfo {author} {\bibfnamefont
  {A.}~\bibnamefont {Peeters}},  \emph {et~al.},\ }\href@noop {} {\bibfield
  {journal} {\bibinfo  {journal} {Nuclear fusion}\ }\textbf {\bibinfo {volume}
  {43}},\ \bibinfo {pages} {1396} (\bibinfo {year} {2003})}\BibitemShut
  {NoStop}%
\bibitem [{\citenamefont {Meyer}\ \emph {et~al.}(2019)\citenamefont {Meyer},
  \citenamefont {Angioni}, \citenamefont {Albert}, \citenamefont {Arden},
  \citenamefont {Parra}, \citenamefont {Asunta}, \citenamefont {De~Baar},
  \citenamefont {Balden}, \citenamefont {Bandaru}, \citenamefont {Behler} \emph
  {et~al.}}]{meyer2019overview}%
  \BibitemOpen
  \bibfield  {author} {\bibinfo {author} {\bibfnamefont {H.}~\bibnamefont
  {Meyer}}, \bibinfo {author} {\bibfnamefont {C.}~\bibnamefont {Angioni}},
  \bibinfo {author} {\bibfnamefont {C.}~\bibnamefont {Albert}}, \bibinfo
  {author} {\bibfnamefont {N.}~\bibnamefont {Arden}}, \bibinfo {author}
  {\bibfnamefont {R.~A.}\ \bibnamefont {Parra}}, \bibinfo {author}
  {\bibfnamefont {O.}~\bibnamefont {Asunta}}, \bibinfo {author} {\bibfnamefont
  {M.}~\bibnamefont {De~Baar}}, \bibinfo {author} {\bibfnamefont
  {M.}~\bibnamefont {Balden}}, \bibinfo {author} {\bibfnamefont
  {V.}~\bibnamefont {Bandaru}}, \bibinfo {author} {\bibfnamefont
  {K.}~\bibnamefont {Behler}},  \emph {et~al.},\ }\href@noop {} {\bibfield
  {journal} {\bibinfo  {journal} {Nucl. Fusion}\ }\textbf {\bibinfo {volume}
  {59}},\ \bibinfo {pages} {112014} (\bibinfo {year} {2019})}\BibitemShut
  {NoStop}%
\end{thebibliography}%

\end{document}